\documentclass[aps,pre,preprint,groupedaddress,showpacs]{revtex4-2}
\usepackage{graphicx}
\usepackage{amssymb}%eu inclui esta
\usepackage{color} %eu inclui esta
\usepackage{epstopdf}%This line makes .eps figures into .pdf - please comment out if not required.

\begin{document}

%\preprint{}

\title{Roles of packing fraction, microscopic friction and projectile spin in cratering by impact\\
	\textnormal{Accepted manuscript for Physical Review E, 105, 034903, (2023), DOI: 10.1103/PhysRevE.107.044901, https://doi.org/10.1103/PhysRevE.107.044901}}



\author{Douglas D. Carvalho}
% \email{douglascarvalho47@gmail.com}
\author{Nicolao C. Lima}
\author{Erick M. Franklin}
 \email{erick.franklin@unicamp.br}
 \thanks{Corresponding author}
\affiliation{%
School of Mechanical Engineering, UNICAMP - University of Campinas,\\
Rua Mendeleyev, 200, Campinas, SP, Brazil\\
 % \textbackslash\textbackslash
}%
%\thanks{}


\date{\today}

\begin{abstract}
From small seeds falling from trees to asteroids colliding with planets and moons, the impact of projectiles onto granular targets occurs in nature at different scales. In this paper, we investigate open questions in the mechanics of granular cratering, in particular the forces acting on the projectile, and the roles of granular packing, grain-grain friction and projectile spin. For that, we carried out DEM (discrete element method) computations of the impact of solid projectiles on a cohesionless granular medium, where we varied the projectile and grain properties (diameter, density, friction and packing fraction) for different available energies (within relatively small values). We found that a denser region forms below the projectile, pushing it back and causing its rebound by the end of its motion, and that solid friction affects considerably the crater morphology. Besides, we show that the penetration length increases with the initial spin of the projectile, and that differences in initial packing fractions can engender the diversity of scaling laws found in the literature. Finally, we propose an \textit{ad hoc} scaling that collapsed our data for the penetration length and can perhaps unify existing correlations. Our results provide new insights into the formation of craters in granular matter.
\end{abstract}

%\pacs{45.70.Qj, 92.40.Pb}
%\keywords{}

\maketitle

%\section{Introduction}

\section{INTRODUCTION}
\label{sec:intro}

The impact of projectiles onto granular targets, with the resulting crater formation and projectile penetration, is frequently observed in nature at different scales. We find it in the collision with the ground of seeds falling from trees, which, when small and light, involves mass, lengths and velocities of the orders of 10 g, 1 cm and 1 cm/s, respectively, corresponding to energy levels of the order of 10$^{-7}$ J and forming cm-size craters. We find it also in the collision of km-size asteroids impacting planets and moons at 10$^{3}$-10$^{4}$ km/h, which corresponds to energies of the orders of 10$^{16}$ (equivalent to a hydrogen bomb) to 10$^{18}$ J, and forms km-size craters. However, the values involved can be much higher: for instance, the \textit{Tycho} and \textit{Posidonius} craters found on Earth's moon have  diameters of approximately 100 km (85 and 95 km, respectively) \cite{Suarez, Kruger}, and the \textit{Odisseus} crater found on Saturn's moon Tethys \cite{Barlow} a diameter of 445 km.

Although the collisional processes are different (for instance, high energies involve melting and evaporation), they bear similarities if we consider the dynamics of the granular material alone, which can be explored if we assure that some dimensionless groups are within certain ranges. It has been shown \cite{Holsapple} that the projectile weight divided by its surface area and normalized by its dynamic pressure is an important dimensionless group for the so-called gravity regime. In this regime, the yield stress of the target is lower than the lithostatic pressure, and the resulting dimensionless number, given by Eq. \ref{Eq:P2}, is the equivalent of the inverse of a Froude number $Fr^{-1}$ (gravitational effects compared to inertia):
	
\begin{equation}
		Fr^{-1} = \frac{D_p g}{V_p^2} \, ,
		\label{Eq:P2}
\end{equation}

\noindent where $D_p$ is the projectile diameter, $V_p$ its velocity at the impact, and $g$ the modulus of gravity acceleration $\vec{g}$. In geophysical problems, typical Froude numbers are within 10$^{-6}$ $\lesssim$ $Fr^{-1}$ $\lesssim$ 10$^{-2}$, where the lower end is in the gravity regime, and the upper end is sometimes acknowledged as being also in that regime \cite{Holsapple, Suarez}. The reason for that uncertainty is the dependence of the impact mechanics on the target material, i.e., the ranges of $Fr^{-1}$ for the gravity regime differ for targets consisting of a continuous material, cohesive grains, or cohesionless grains. As pointed out by Holsapple \cite{Holsapple}, the range of $Fr^{-1}$ for the gravity regime is larger when the target consists of cohesionless grains, so that $Fr^{-1}$ $\geq$ 10$^{-2}$ is sometimes considered in that regime.

Concerning the projectile, its dynamics is subjected to a deceleration once the impact takes place \cite{Poncelet, Tsimring, Katsuragi, Suarez, Katsuragi3}. In general, it has two distinct phases \cite{Goldman, Katsuragi}: In the first phase, the grains in the impact region are fluidized and the projectile penetrates the target with a predominant inertial drag, while in the second phase the bed hardens again and the projectile continues its penetration with a depth-dependent frictional drag \cite{Goldman}. The projectile thus decelerates while penetrating the granular medium because of the opposing drag and, just before reaching a full stop, suffers a discontinuity in its acceleration: the dynamic drag is changed to a static force that supports the projectile \cite{Goldman, Katsuragi}. For a vertical coordinate $y$ oriented downwards and a force drag $\vec{F}_{drag}$ oriented upwards (with respect to gravity acceleration $\vec{g}$), the resultant force $\vec{F_p}$ (oriented upwards) acting on a solid projectile of mass $m_p$ while it penetrates the granular bed is given by Eq. (\ref{Eq:intruder}), 

\begin{equation}
	F_p = m_p\frac{d^2y}{dt^2} = - m_p g + F_{drag} \, ,
	\label{Eq:intruder}
\end{equation}

\noindent where $F_p$ and $F_{drag}$ are the moduli of $\vec{F_p}$ and $\vec{F}_{drag}$, respectively, $V$ = $dy/dt$ is the instantaneous velocity of the projectile, and $F_{drag}$ = $\xi V^2 + \kappa y$ (inertial and friction terms), $\xi$ and $\kappa$ being parameters that depend on the projectile characteristics (density and shape) \cite{Katsuragi, Pacheco}. Umbanhowar and Goldman \cite{Umbanhowar} proposed that Eq. (\ref{Eq:intruder}) is incomplete to describe projectiles impacting targets with different packing fractions $\phi$, being valid only close to a critical packing $\phi_{cps}$.

Goldman and Umbanhowar \cite{Goldman} observed fluctuations during the inertial phase, and conjectured that such fluctuations are due to the formation and collapse of granular chains. They showed that, at the acceleration discontinuity that occurs at the end of the motion, the projectile moves upwards before reaching full stop, and also observed that the impact time $t_c$ (time interval from reaching the target to full stop) is approximately independent of $V_p$ above a threshold value. Finally, they proposed that in Eq. (\ref{Eq:intruder}) the inertial term dominates at high velocity and shallow penetration, while at low velocities and deep penetrations a viscous-like term linear in $V$ must be added and dominates the drag force together with the frictional term.

Concerning the morphology, the scales for the diameter $D_c$ and depth $h_c$ of craters can be obtained from physical arguments \cite{Amato, Uehara, Uehara2, Walsh}. For the gravity regime, it is expected that the available energy at the impact $E$ is dissipated by excavating the crater, i.e., displacing the crater volume ($\sim D_c^3$) by a distance proportional to $h_c$. By hypothesizing that $h_c \sim D_c$ in this regime, we find $D_c \sim E^{1/4}$ and, therefore, $h_c \sim E^{1/4}$. However, more sophisticated computations and experiments were carried out to better understand the physical mechanisms involved in the gravity regime. The crater diameter $D_c$ has been reported \cite{Uehara, Uehara2, Walsh, deVet} to, indeed, scale as $D_c \sim E^{1/4}$, but different scalings were obtained for $h_c$. For the latter, some authors found that $h_c \sim D_c$ and then $h_c \sim E^{1/4}$ \cite{Walsh, deVet}, but others found different scales, such as $h_c \sim V_p^{2/3}$ (Uehara et al. \cite{Uehara, Uehara2}, where in their case $h_c$ was equal to the depth $\delta$ reached by the projectile).

In particular, Ciamarra et al. \cite{Ciamarra} investigated experimentally and numerically the impact of a projectile onto a two-dimensional granular medium (disks), and found that $t_c$ is independent of $V_p$, so that the projectile penetration $\delta$ depends on the impact velocity. They found a constant deceleration that is proportional to $V_p$, which explains the independence of $t_c$, but that is in disagreement with the direct measurements made later by Goldman and Umbanhowar \cite{Goldman}. Uehara et al. \cite{Uehara, Uehara2} released spheres of different densities $\rho_p$ from different initial heights $h$ onto cohesionless beads and, for partially penetrating spheres ($\delta$ = $h_c$), found that $D_c$ $\sim$  $\left(\rho_p D_p^3 H \right)^{1/4}$ $\sim$ $E^{1/4}$ (according to predictions), where $H$ is the total drop distance (vertical distance traveled by the projectile, including the penetration depth $\delta$, so that $H$ = $h + \delta$). However, they found that $h_c \sim H^{1/3}$, $h_c$ not scaling with $E$, so that the crater aspect ratio is not necessarily fixed. They also found that the friction and restitution coefficients of the projectile and diameter of the grains do not affect the crater morphology. On the other hand, de Bruyn and Walsh \cite{deBruyn} found experimentally that $\delta$ $\sim$ $V_p$ and, by varying the packing fraction $\phi$, that $\delta$ $\sim$ $\phi$. They checked the dependency on $\phi$ against the model of Uehara et al. \cite{Uehara, Uehara2}, but the existence of deviations made them propose a different correlation with $h$ and $D_p$. Until now, the scaling laws for the crater $h_c$ and penetration $\delta$ depths remain without a consensus.

Another important question is how the microscopic friction influences cratering. Tsimring and Volfson \cite{Tsimring} proposed that the microscopic friction dissipates a significant part of the available energy (they found approximately 70\% of the impact energy), which was later corroborated by the 2D (two-dimensional) DEM (discrete element method) simulations of Kondic et al. \cite{Kondic}. However, Seguin et al. \cite{Seguin} found the contrary: that impact on frictionless grains causes roughly the same penetration depths and stopping times. According to the authors, the inelastic collisions would be the main responsible for the energy dissipation, the microscopic friction engendering only minor effects. One possible explanation for an independence on the solid friction was pointed out by Suarez \cite{Suarez}: in quasi-static motion (existing far from the projectile), grains would by turns form and break granular chains. This leads to jammed and unjammed states occurring in grain-grain interfaces (but not in the projectile-grain interface), so that compressive stresses within grains would be more important than shear stresses. The independence on the microscopic friction is, however, still object of debate.

Although considerable progress on the mechanics of impacts and crater formation was made from previous studies, many questions remain open, such as the scaling laws for the penetration depth, and the roles of friction and initial packing fractions. Other questions are still to be investigated, such as how the initial spin of projectiles (rotational kinetic energy) affects cratering. In this paper, we inquire into those questions by carrying out DEM computations of the impact of solid projectiles onto a cohesionless granular medium (in the gravity regime). For different projectile and grain properties (diameter, density, friction coefficients and packing fraction), we measured the morphology of craters, fluctuations of grains, and resultant force on the projectile. We show that the scales of craters and the dynamics of projectiles compare well with some of the existing scaling laws, but not with others. We find that, after an initial fluidization, a denser region forms below the projectile, which pushes it back and causes its rebound by the end of its motion, and that solid friction affects considerably the crater morphology. In addition, we show that the penetration length $\delta$ increases with the initial spin (angular velocity) of the projectile and that differences in the initial packing fraction $\phi$ engender the diversity of scaling laws found in the literature. Finally, we propose an \textit{ad hoc} scaling for $\delta$ involving $\phi$ that can, perhaps, unify the existing correlations. Our results provide new insights into the formation of craters by the impact of solid projectiles.

\section{MODEL DESCRIPTION AND NUMERICAL SETUP}
\label{sec:mumerical}

We carried out 3D DEM computations \cite{Cundall}, where the Newton's law of motion is computed for each individual particle by using the open-source code LIGGGHTS \cite{Kloss, Berger}. Basically, the code computes the linear (Eq. (\ref{Fp})) and angular (Eq. (\ref{Tp})) momentum equations at each time step,

\begin{equation}
	m\frac{d\vec{u}}{dt}= \vec{F}_{c} + m\vec{g} \, ,
	\label{Fp}
\end{equation}

\begin{equation}
	I\frac{d\vec{\omega}}{dt}=\vec{T}_{c} \, ,
	\label{Tp}
\end{equation}

\noindent where, for each particle, $m$ is the mass, $\vec{u}$ is the velocity, $I$ is the moment of inertia, $\vec{\omega}$ is the angular velocity, $\vec{F}_{c}$ is the resultant of contact forces between solids (Eq. (\ref{Fc})), and $\vec{T}_{c}$ is the resultant of contact torques between solids (Eq. (\ref{Tc})).

\begin{equation}
	\vec{F}_{c} = \sum_{i\neq j}^{N_c} \left(\vec{F}_{c,ij} \right) + \sum_{i}^{N_w} \left( \vec{F}_{c,iw} \right)
	\label{Fc}
\end{equation}

\begin{equation}
	\vec{T}_{c} = \sum_{i\neq j}^{N_c} \vec{T}_{c,ij} + \sum_{i}^{N_w} \vec{T}_{c,iw}
	\label{Tc}
\end{equation}

In Eqs. (\ref{Fc}) and (\ref{Tc}), $\vec{F}_{c,ij}$ and $\vec{F}_{c,iw}$ are the contact forces between particles $i$ and $j$ and between particle $i$ and the wall, respectively, $\vec{T}_{c,ij}$ and $\vec{T}_{c,iw}$ are torques due to the tangential component of the contact forces between particles $i$ and $j$ and between particle $i$ and the wall (both considering rolling resistance), respectively, $N_c$ - 1 is the number of particles in contact with particle $i$, and $N_w$ is the number of particles in contact with the wall. The contact forces ($\vec{F}_{c,ij}$ and $\vec{F}_{c,iw}$) are computed using the elastic Hertz-Mindlin contact model \cite{direnzo}, described in Appendix \ref{appendix}. In the contact torques ($\vec{T}_{c,ij}$ and $\vec{T}_{c,iw}$), the rolling resistance is considered through a coefficient of rolling friction $\mu_r$, also described in Appendix \ref{appendix}. The torque due to rolling resistance is important if angular grains (sand, for example) are modeled as spherical particles with the angularity effects embedded in the rolling friction \cite{Derakhshani}. Such effects are negligible for perfect spherical grains.

The computed system consisted of $N$ $\sim$ 10$^6$ spheres with diameter $d$ and density $\rho$, forming a granular bed in a cylindrical container, and a projectile with diameter $D_p$ and density $\rho_p$. Prior to each simulation, around 10$^6$ spheres with a Gaussian distribution for $d$ were randomly arranged in space, and let to fall freely in the container and settle until a low level of kinetic energy was attained. By varying the initial value of the grain-grain friction coefficient $\mu_{gg}$, we obtained different packing fractions $\phi$ for the bed, after which we changed $\mu_{gg}$ back to the correct value. The grains were then allowed to relax, and only afterward the simulations began. The distribution of diameters used in the simulations are shown in Tab. \ref{tab_diameters}. We then computed the minimum height necessary for having a horizontal surface and deleted all the grains above that height (around 10$^4$ grains removed). Depending on the properties of the spheres and their initial number, the number $N$ that remained in the computational domain varied. The granular beds had a diameter $D_{bed}$ = 125 mm and heights $h_{bed}$ = 67.0-76.5 mm (depending on the packing fraction). In order to avoid strong confinement effects, the bed dimensions are equal to the largest dimensions investigated by Seguin et al. \cite{Seguin2}.

\begin{table}[!h]
	\centering
	\caption{Distribution of diameters for the settling grains: number of grains $N_d$ for each diameter $d$.}
	\label{tab_diameters}
	\begin{tabular}{|c|c|c|c|c|c|}
		\hline
		$d$ (mm) & 0.6 & 0.8 & 1.0 & 1.2 & 1.4\\
		\hline
		$N_d$ ($\phi$ = 0.554)  & 21524 & 128125 & 643002 & 128053 & 21421\\
		\hline
		$N_d$ ($\phi$ = 0.575-0.632)  & 21483 & 128214 & 642847 & 127831 & 21340\\
		\hline
	\end{tabular}
\end{table}

\begin{table}[!h]
	\centering
	\caption{Properties of materials used in the simulations: $E$ is Young's modulus, $\nu$ is the Poisson ratio, and $\rho$ is the material density. The last column corresponds to the diameter of the considered object.}
	\label{tabmaterials}
	\begin{tabular}{l|c|c|c|c|c}
		\hline
		& \textbf{Material} & \textbf{$E$ (Pa)} & \textbf{$\nu$} & \textbf{$\rho$ (kg/m$^{3}$)}& \textbf{Diameters (mm)}\\
		\hline
		Projectile & Steel\footnotesize{$^{(1)}$} & $0.2 \times 10^{11}$  & 0.3 & 7865 & 15\\
		Grains & Sand\footnotesize{$^{(1)-(2)}$} & $0.1 \times 10^{9}$ & 0.3 & 2600 & 0.6 $\leq$ $d$ $\leq$ 1.4 \\
		Walls & Steel \footnotesize{$^{(1)}$} & $0.2 \times 10^{12}$ & 0.3 & 7865 & 125\\    
		\hline
		\multicolumn{3}{l}{\footnotesize{$^{(1)}$ Ucgul et al. \cite{Ucgul1, Ucgul2, Ucgul3}}} \\
		\multicolumn{3}{l}{\footnotesize{$^{(2)}$ Derakhshani et al. \cite{Derakhshani}}}\\		
	\end{tabular}
\end{table}

\begin{figure}[h!]
	\begin{center}
		\includegraphics[width=.5\linewidth]{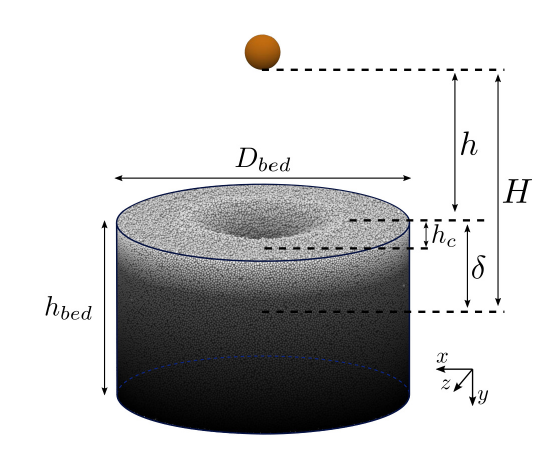}\\
	\end{center}
	\caption{Layout of the numerical setup. The origin of the coordinate system is on the bed surface, in the center of the domain; however, it is shown on the bottom right for better visualization.}
	\label{fig:setup}
\end{figure}

At the beginning of the simulations, the projectile is put into motion in order to collide with the granular bed with collision velocities $V_p$ that are related with the free-fall height $h$ (distance from the bed surface to the initial position of the projectile centroid minus its radius, $V_p$ = $\sqrt{2gh}$). With that, Froude numbers were within 3.75 $\times$ 10$^{-3}$ $\leq$ $Fr^{-1}$ $\leq$ 3, all of which we consider in the gravity regime. Figure \ref{fig:setup} shows a layout of the numerical setup, and animations showing impacts and cratering are available in the Supplemental Material \cite{Supplemental}.

\begin{table}[!h]
	\centering
	\caption{Coefficients used in the numerical simulations.}
	\label{tabcoefficients}
	\begin{tabular}{l|c|c}
		\hline
		\textbf{Coefficient}  & \textbf{Symbol} & \textbf{Value} \\
		\hline		  
		Restitution coefficient (grain-grain)\footnotesize{$^{(1)}$} & $\epsilon_{gg}$ & 0.6 \\
		Restitution coefficient (grain-projectile)\footnotesize{$^{(1)}$} & $\epsilon_{gp}$ & 0.6 \\
		Restitution coefficient (grain-wall)\footnotesize{$^{(1)}$} & $\epsilon_{gw}$ & 0.6 \\
		Fiction coefficient (grain-grain)\footnotesize{$^{(1)-(2)}$} & $\mu_{gg}$ & 0.52 \\
		Friction coefficient (grain-projectile)\footnotesize{$^{(1)}$} & $\mu_{gp}$ & 0.5 \\
		Friction coefficient (grain-walls)\footnotesize{$^{(1)}$} & $\mu_{gw}$ & 0.5 \\		
		Coefficient of rolling friction (grain-grain)\footnotesize{$^{(2)}$} & $\mu_{r,gg}$ & 0.3\\
		Coefficient of rolling friction (grain-projectile)\footnotesize{$^{(1)}$} & $\mu_{r,gp}$ & 0.05\\
		Coefficient of rolling friction (grain-wall)\footnotesize{$^{(1)}$} & $\mu_{r,gw}$ & 0.05\\
		\hline
		\multicolumn{3}{l}{\footnotesize{$^{(1)}$ Ucgul et al. \cite{Ucgul1, Ucgul2, Ucgul3}}} \\
		\multicolumn{3}{l}{\footnotesize{$^{(2)}$ Derakhshani et al. \cite{Derakhshani}}}
	\end{tabular}
\end{table}

We used different properties for the grains and projectile, listed in Tabs. \ref{tabmaterials} and \ref{tabcoefficients} together with those for walls. We used the real Young's modulus $E$, with the exception of projectiles in steel, for which we used a value that was smaller by one order of magnitude. Because steel has the higher Young's modulus among the used materials, and since the projectile suffers a considerable number of energetic impacts (much larger than the walls), this numerical artifice increased the necessary time step without affecting significantly the results \cite{Lommen}. In our simulations, all the coefficients were taken from the literature, and the sand grains were modeled as spherical particles with angularity effects embedded in the rolling friction, for which we used the value $\mu_r$ = 0.3 validated by Derakhshani et al. \cite{Derakhshani} (we validated the friction coefficients listed in Tab. \ref{tabcoefficients} by measuring the angles of repose obtained numerically, see the Supplemental Material \cite{Supplemental} for details). Although we present results for fixed $\rho_p$ and $D_p$ (as listed in Tab. \ref{tabmaterials}) in the following, we carried out simulations with 2685 kg/m$^3$ $\leq$ $\rho_p$ $\leq$ 11865 kg/m$^3$ and 5 mm $\leq$ $D_p$ $\leq$ 30 mm (results available in the Supplemental Material \cite{Supplemental}).

We used a time step $\Delta t = 8 \times 10^{-7}$ s in our computations, which assured $\Delta t$ less than 10 \% of the Rayleigh time \cite{Derakhshani} for all particles. More details about the numerical setup are available in the Supplemental Material \cite{Supplemental} and in an open repository \cite{Supplemental2}.

\section{\label{sec:Res} RESULTS AND DISCUSSION}

\subsection{\label{sec:morphology} Morphology of craters}

\begin{figure}[h!]
	\begin{center}
		\begin{minipage}{0.49\linewidth}
			\begin{tabular}{c}
				\includegraphics[width=0.80\linewidth]{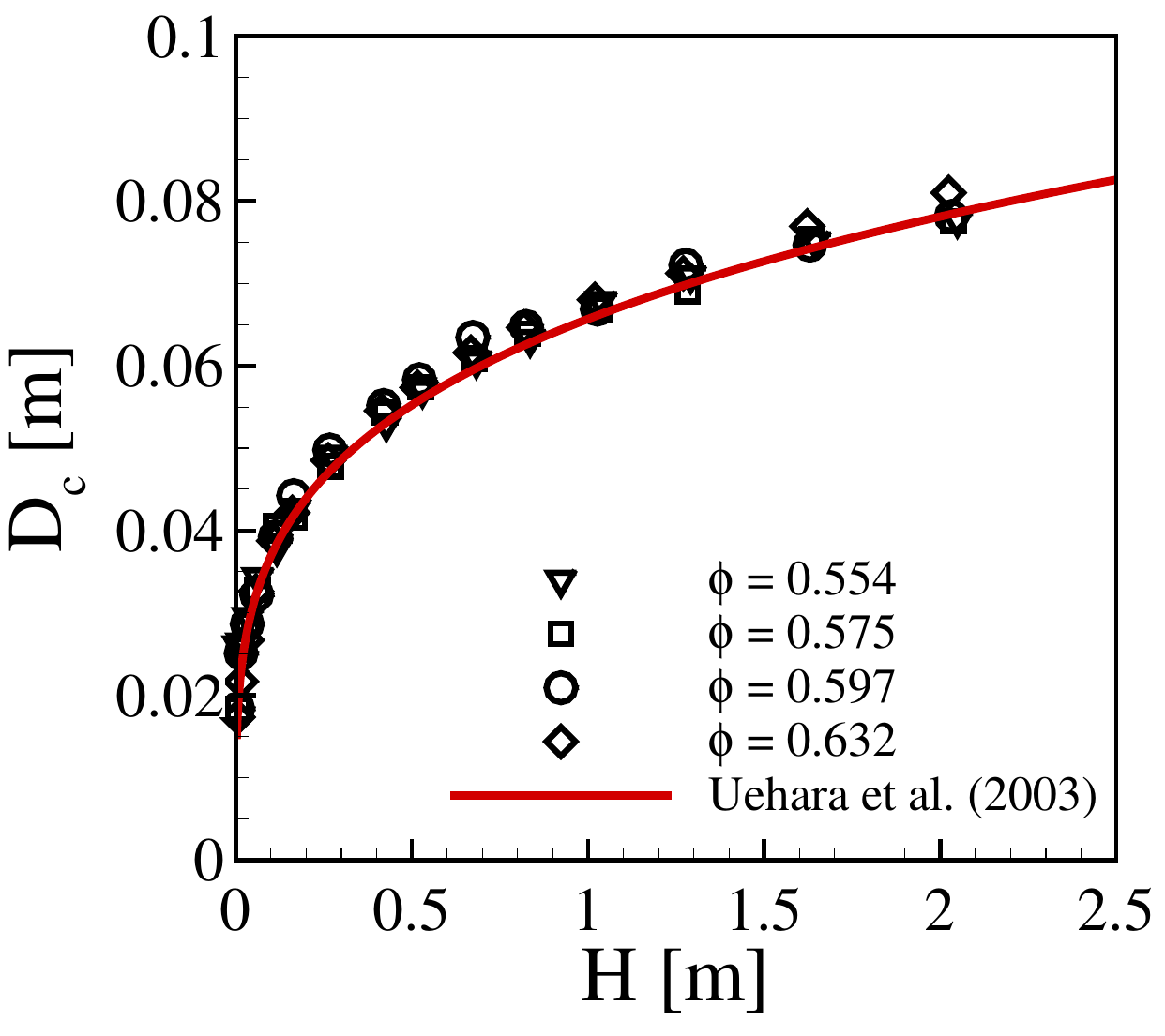}\\
				(a)
			\end{tabular}
		\end{minipage}
		\hfill
		\begin{minipage}{0.49\linewidth}
			\begin{tabular}{c}
				\includegraphics[width=0.80\linewidth]{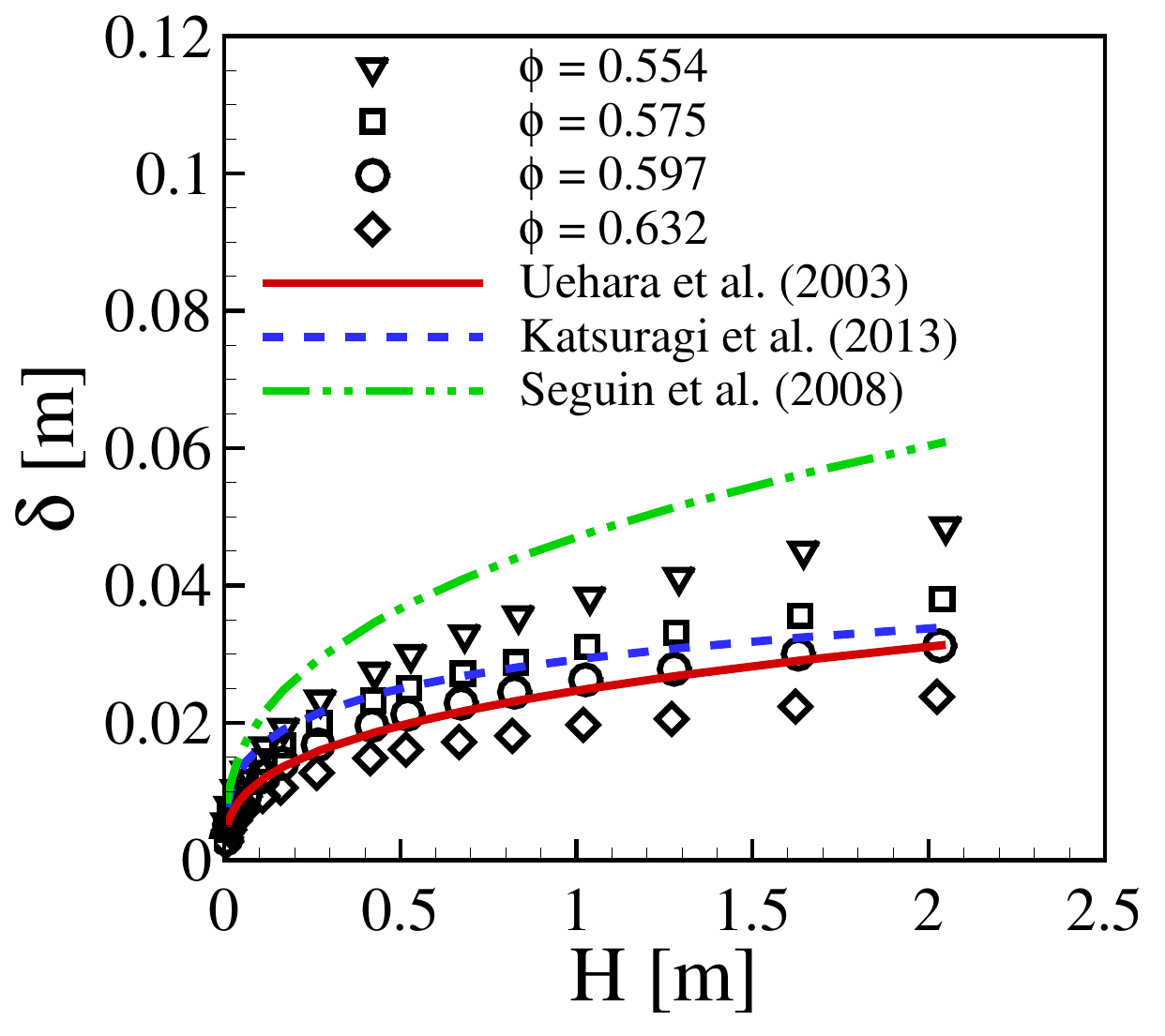}\\
				(b)
			\end{tabular}
		\end{minipage}
		\hfill
		\begin{minipage}{0.49\linewidth}
			\begin{tabular}{c}
				\includegraphics[width=0.80\linewidth]{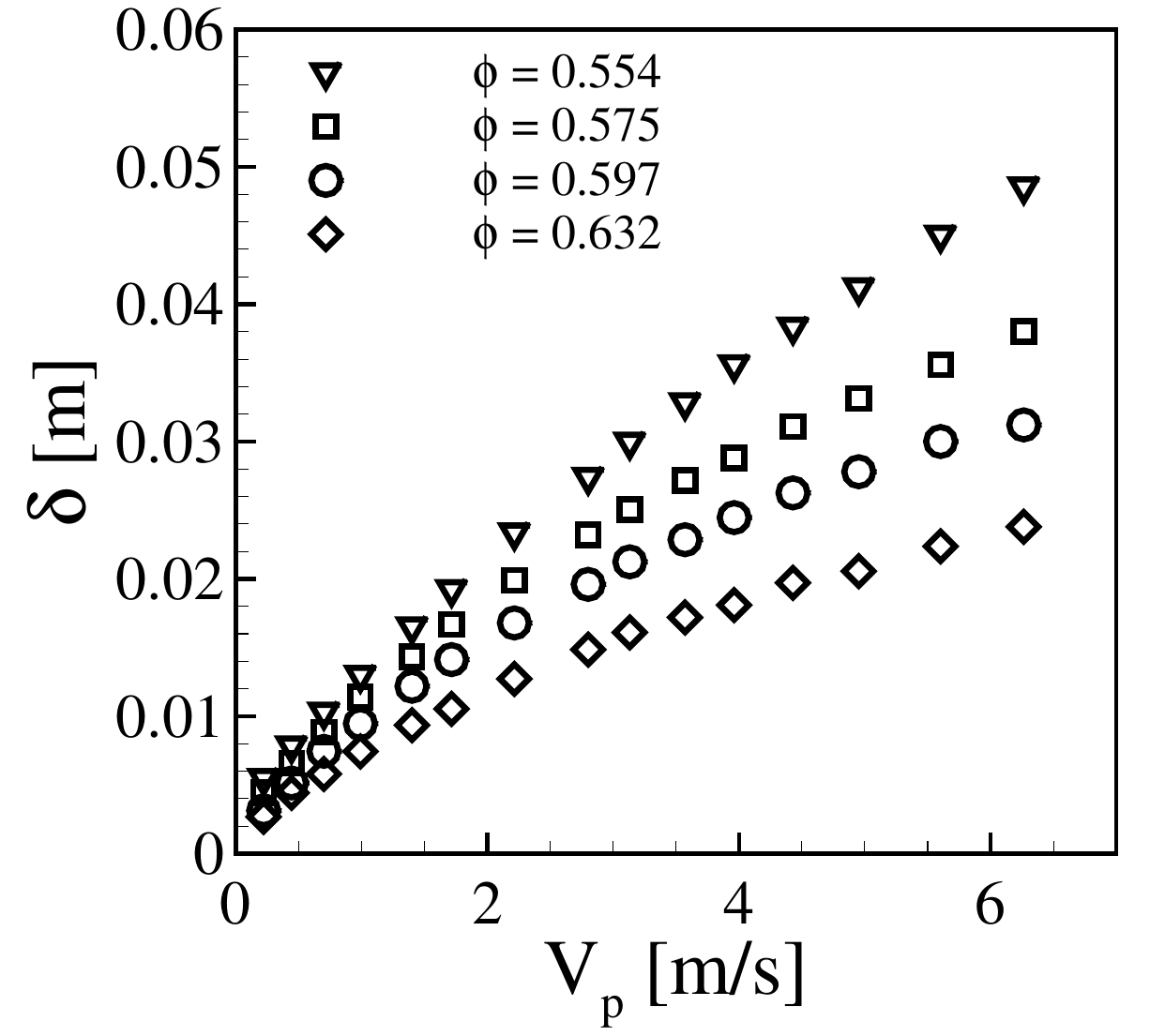}\\
				(c)
			\end{tabular}
		\end{minipage}
		\hfill
	\end{center}
	\caption{Morphological aspects: (a) Crater diameter $D_c$ as a function of the total drop distance $H$; (b) depth $\delta$ reached by the projectile as a function of $H$; and (c) $\delta$ as a function of the projectile velocity at the impact $V_p$. In Figs. (a) and (b), the corresponding correlations proposed by Uehara et al. \cite{Uehara}, Katsuragi et al. \cite{Katsuragi2} and Seguin et al. \cite{Seguin2} are also plotted, and Figs (a) to (c) are parameterized by the initial packing fraction.}
	\label{fig:morphology}
\end{figure}

Although extensively investigated over the last decades, the morphological laws for craters are still object of debate (with the exception, perhaps, of $D_c$), and correlations available in the literature often mix data obtained under different (if not unknown) packing fractions. In other instances, functional relations are based on different parameters ($h$ instead of $H$, for example). Therefore, we investigate initially how the crater diameter $D_c$ and the depth  $\delta$ reached by the projectile behave with varying the drop distance $H$ (or the height to impact $h$, related to the velocity at the impact $V_p$), and compare them with morphological laws found in the literature. In addition, different from previous studies, we evaluate how those relations vary with the initial packing fraction $\phi$ of the bed.

Figure \ref{fig:morphology}(a) shows the crater diameter $D_c$ as a function of the total drop distance $H$, parameterized by the initial packing fraction $\phi$, and the correlation proposed by Uehara et al. \cite{Uehara}, given by Eq. (\ref{Dc_uehara}),

\begin{equation}
	D_{c} = 0.90\bigg(\frac{\rho_{p}}{\rho \mu_{rep}^2}\bigg)^{1/4}D_{p}^{3/4}H^{1/4} \,,
	\label{Dc_uehara}
\end{equation} 

\noindent where $\mu_{rep}$ is the macroscopic friction measured as the tangent of the angle of repose, which Uehara et al. \cite{Uehara} considered equal to $\mu_{gg}$. In fact, we find a consensus in the literature that $D_c$ varies with $H^{1/4}$, and our data shows the same, with a collapse of data for all the different packing fractions used in the simulations. Therefore, $D_c$ is independent of $\phi$, and this is the main reason for the existing consensus since the different experiments reported in the literature were conducted at different packing fractions. The same does not occur with the depth $\delta$ reached by the projectile. Figure \ref{fig:morphology}(b) shows $\delta$ as a function of $H$, parameterized by $\phi$, and the  corresponding correlations proposed by Uehara et al. \cite{Uehara},  Katsuragi et al. \cite{Katsuragi2} and Seguin et al. \cite{Seguin2}, given by Eqs. (\ref{eq_uehara}), (\ref{eq_katsuragi}) and (\ref{eq_seguin}), respectively. We observe a clear dependence of $\delta$ on $\phi$, and that correlations give different results. The discrepancies between the existing correlations are thus, at least in part, due to the different packing fractions of the experiments they came from. The dependence of $\delta$ on $\phi$ is shown also in Fig. \ref{fig:morphology}(c) in terms of the projectile velocity at the impact $V_p$. We observe that the data diverge for increasing values of $V_p$, presenting a non-linear variation with $V_p$ for higher values of $\phi$. This is in contrast with Katsuragi and Durian \cite{Katsuragi} and Goldman and Umbanhowar \cite{Goldman}, who found that $\delta$ varies linearly with $V_p$. However, we note that: (i) most of the data presented by Refs. \cite{Katsuragi, Goldman} are within 0 m/s $\leq$ $V_p$ $\leq$ 4 m/s, for which the dependencies tend to appear more linear; and (ii) we controlled the packing fraction in each of our simulations (different from previous works), finding considerable deviations for higher values of $\phi$. If we consider 0 m/s $\leq$ $V_p$ $\leq$ 4 m/s, the curves in Fig. \ref{fig:morphology}(c) become roughly linear. The correlations plotted in Fig. \ref{fig:morphology}(b) are presented below.

\noindent (i) Correlation proposed by Uehara et al. \cite{Uehara}:

\begin{equation}
	\delta = 0.14\bigg(\frac{\rho_{p}}{\rho\mu_{rep}^2}\bigg)^{1/2}D_{p}^{2/3}H^{1/3}
	\label{eq_uehara}
\end{equation}

\noindent (ii) Correlation proposed by Katsuragi et al. \cite{Katsuragi2}:

\begin{equation}
	\frac{2\delta}{d_{1}} = 1 + \frac{2m_p g}{\kappa d_1} + \mathcal{W}\Bigg(\frac{2m _p V_{p}^{2} - 2m_p gd_{1}-\kappa d_{1}^{2}}{\kappa d_{1}^{2} e^{1+2m_p g/\kappa d_{1}}}\Bigg) \,,
	\label{eq_katsuragi}
\end{equation}

\noindent where $\mathcal{W}(x)$ is the Lambert function, and $\kappa$ and $d_{1}$ are constants given by Eq. (\ref{eq_katsuragi2}):

\begin{equation}
	\frac{d_{1}}{D_{p}} = \Bigg(\frac{0.25}{\mu_{rep}} \Bigg)\Bigg(\frac{\rho_{p}}{\rho}\Bigg); \qquad \frac{\kappa D_{p}}{m_p g} = 12\mu_{rep}\Bigg(\frac{\rho}{\rho_{p}}\Bigg)^{\frac{1}{2}}
	\label{eq_katsuragi2}
\end{equation}

\noindent OBS: in Eq. (\ref{Eq:intruder}), $\xi$ = $m_p/d_1$.

\noindent (iii) Correlation proposed by Seguin et al. \cite{Seguin2}: 

\begin{equation}
	\frac{\delta}{D_{p}} = A\Bigg(\frac{\rho_{p}}{\rho}\Bigg)^{\beta}\Bigg(\frac{H}{D_{p}}\Bigg)^{\lambda} \,,
	\label{eq_seguin}
\end{equation}

\noindent where $A$ = 0.37 $\pm$ 0.01, $\beta$ = 0.61 $\pm$ 0.02 and  $\lambda$ = 0.40 $\pm$ 0.04. We note that in Fig. \ref{fig:morphology}(b) we used the lower limit of these constants.

Unlike most of previous experiments, de Bruyn and Walsh \cite{deBruyn} varied the packing fraction and, by modeling the granular system as a Bingham fluid, found that $\delta$ $\sim$ $\phi$. They proposed a correlation where $\delta$ $\sim$ $h^{ 1/2} D_p^{1/2}$, which contrasts with the above ones (Eqs. \ref{eq_uehara} to \ref{eq_seguin}). However, as pointed out by the authors, they expected that inaccuracies in their measurements of $\phi$ could affect the results. From our numerical data, we also noticed that the penetration depth depends on the packing fraction, producing thus different correlations for $\delta$ with $H$ or $V_p$. We propose an \textit{ad hoc} scaling that collapses our data, but without additional modeling (we maintain the discrete nature of granular matter in our analysis). Our objective in proposing this \textit{ad hoc} scaling is simply to collapse our $\delta(H)$ data for different values of $\phi$, showing that, perhaps, the existing correlations can be unified by considering a dependency on $\phi$.

\begin{figure}[h!]
	\begin{center}
		\begin{minipage}{0.49\linewidth}
			\begin{tabular}{c}
				\includegraphics[width=0.80\linewidth]{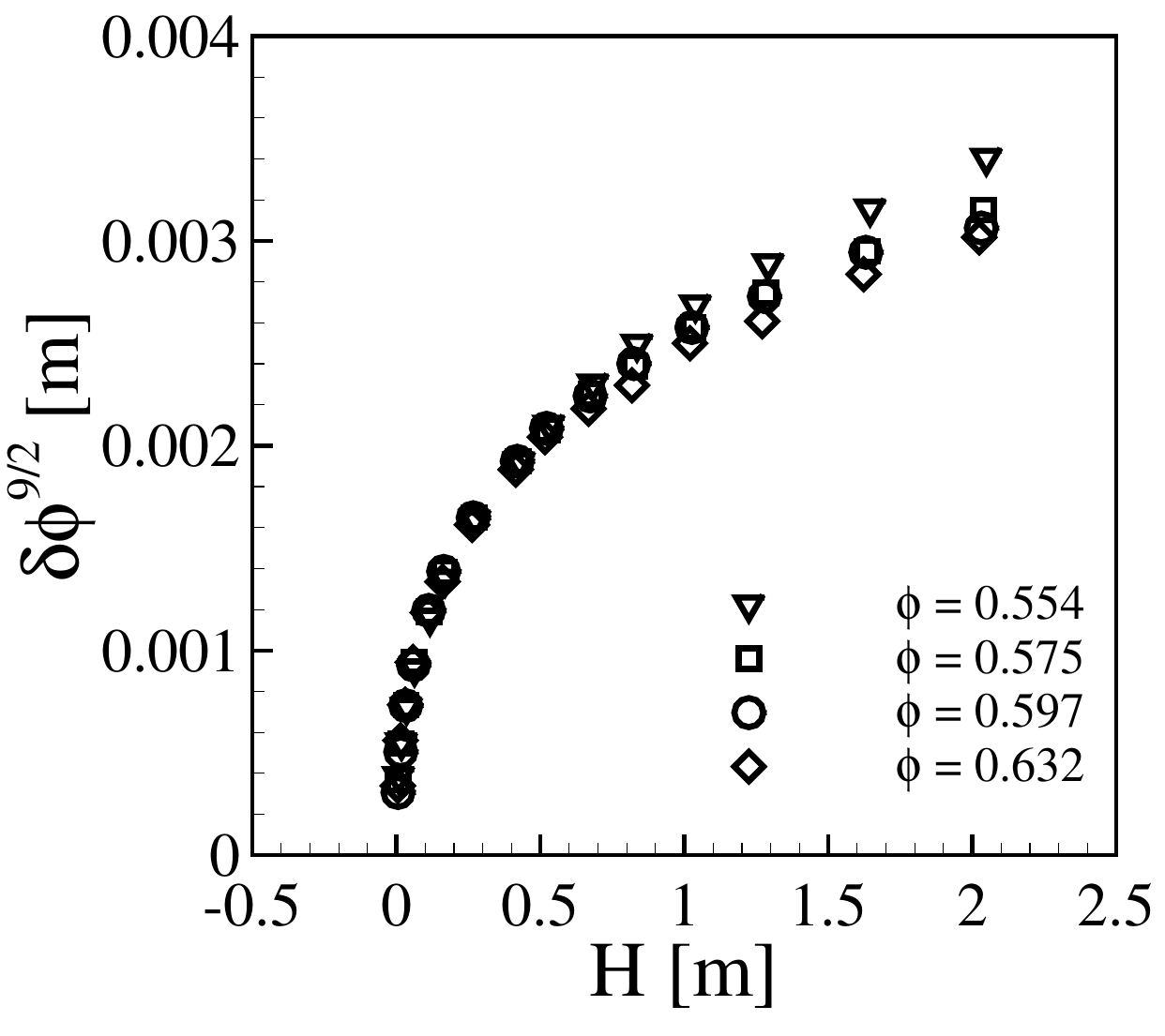}\\
				(a)
			\end{tabular}
		\end{minipage}
		\hfill
		\begin{minipage}{0.49\linewidth}
			\begin{tabular}{c}
				\includegraphics[width=0.80\linewidth]{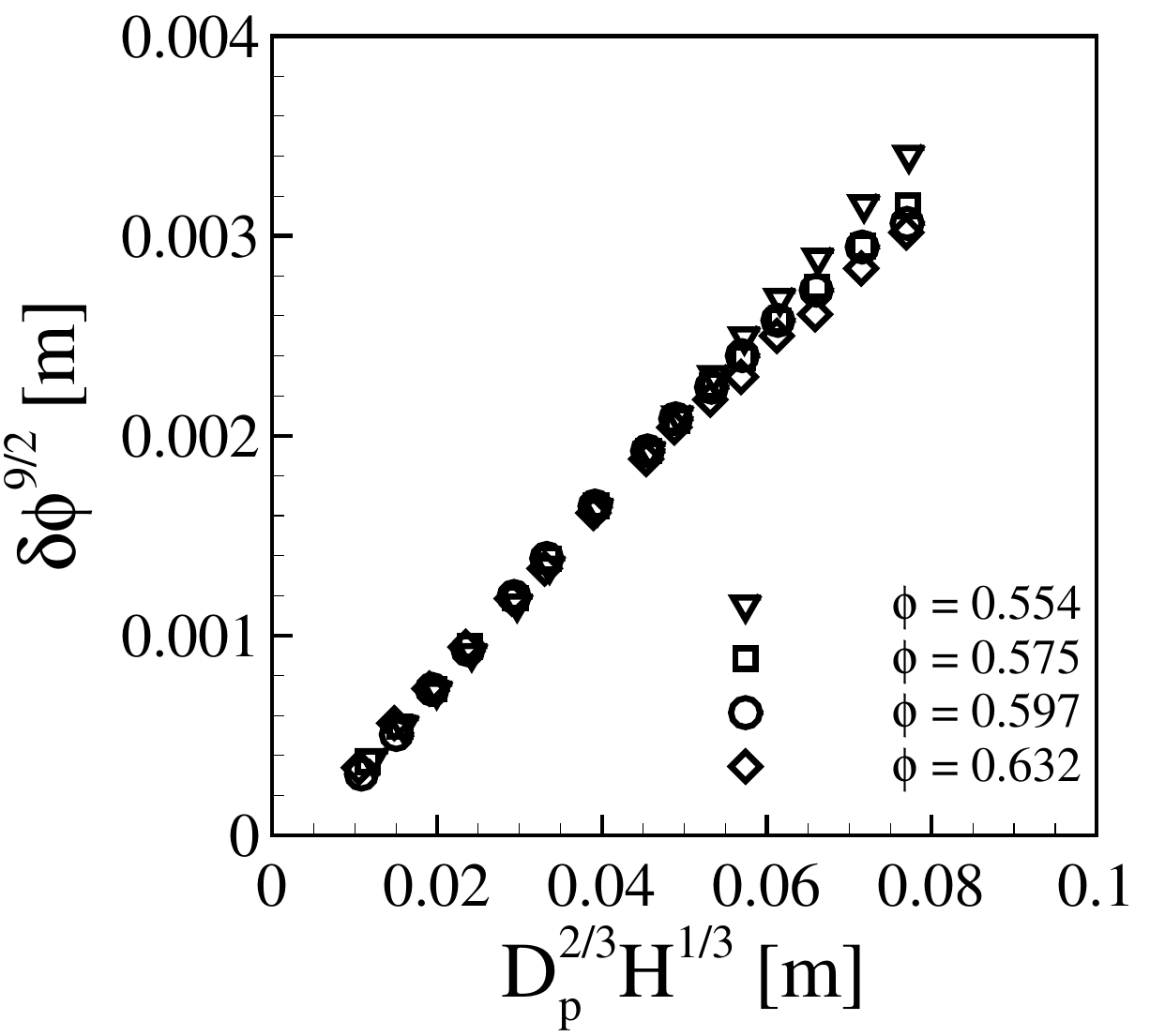}\\
				(b)
			\end{tabular}
		\end{minipage}
		\hfill
	\end{center}
	\caption{Depth reached by the projectile multiplied by a power of the packing fraction, $\delta \phi^{9/2}$, as a function of (a) the drop distance $H$, and (b) $D_{p}^{2/3}H^{1/3}$, parameterized by $\phi$.}
	\label{fig:morphology2}
\end{figure}

Figure \ref{fig:morphology2}(a) shows $\delta(H)$ for our simulations, where a factor  $\phi^{9/2}$ was introduced in order to collapse the data into a master curve. We notice that the collapse is reasonable, indicating that $\phi$ is a parameter to be taken into account. By considering specifically the correlation proposed by Uehara et al. \cite{Uehara} (Eq. (\ref{eq_uehara})), Fig. \ref{fig:morphology2}(b) shows $\delta\phi^{9/2}$ as a function of $D_{p}^{2/3}H^{1/3}$. The data collapse and follow a master line, with some dispersion for higher values of $H$. This indicates that by taking into account a term $\phi^{n}$ (as in Eq. (\ref{proposed_fitting})), where $n$ is a coefficient, some of the existing expressions may be turned into universal correlations. For our data, $n$ = 9/2 is a reasonable value (please note that $n$ = 4.74 gave a slightly better collapse than 9/2).

\begin{equation}
	\delta\phi^{n} \sim D_{p}^{2/3}H^{1/3}
	\label{proposed_fitting}
\end{equation}

\subsection{\label{sec:drag} Forces on the projectile and stopping time}

\begin{figure}[h!]
	\begin{center}
		\begin{minipage}{0.49\linewidth}
			\begin{tabular}{c}
				\includegraphics[width=0.8\linewidth]{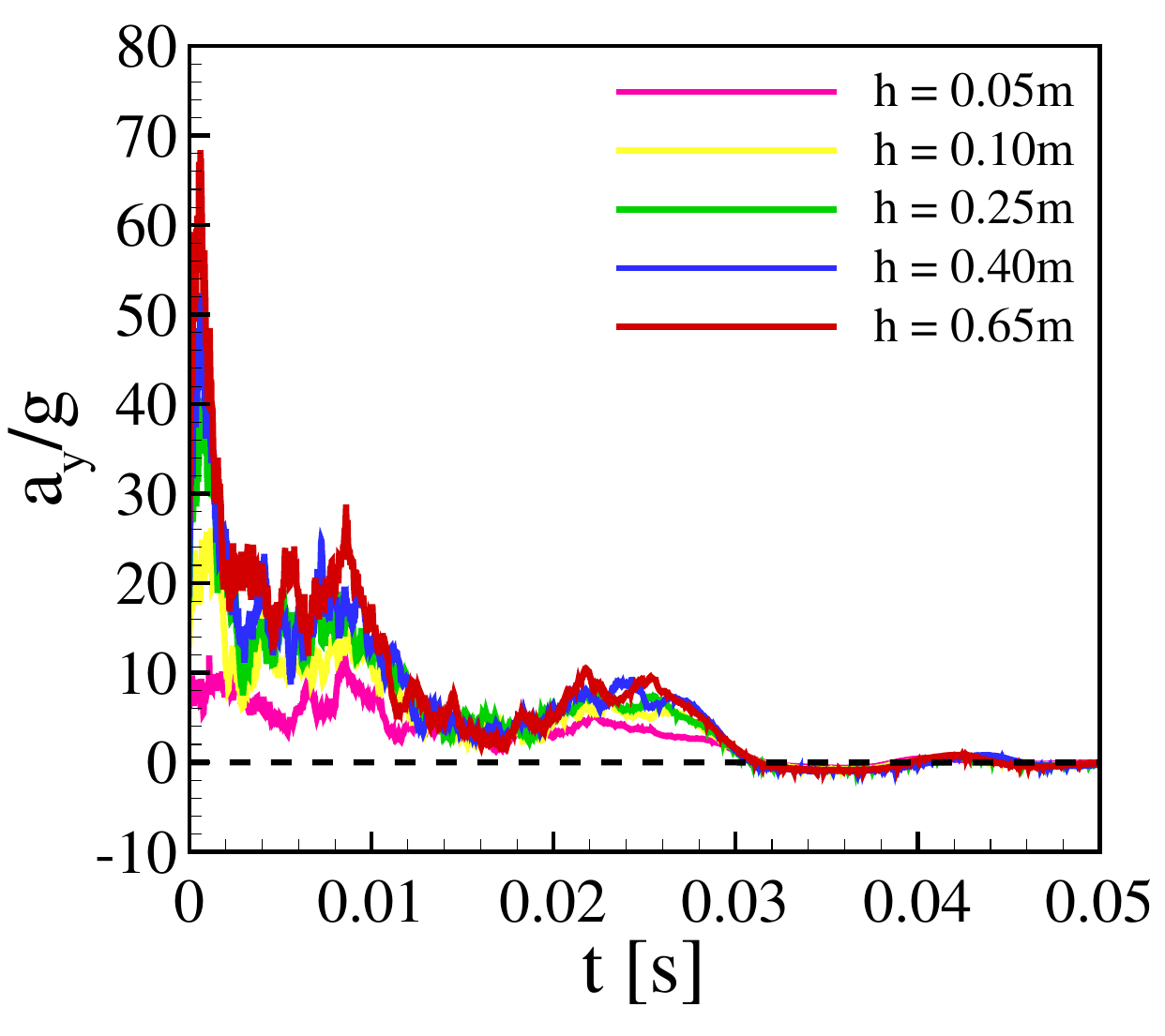}\\
				(a)
			\end{tabular}
		\end{minipage}
		\hfill
		\begin{minipage}{0.49\linewidth}
			\begin{tabular}{c}
				\includegraphics[width=0.8\linewidth]{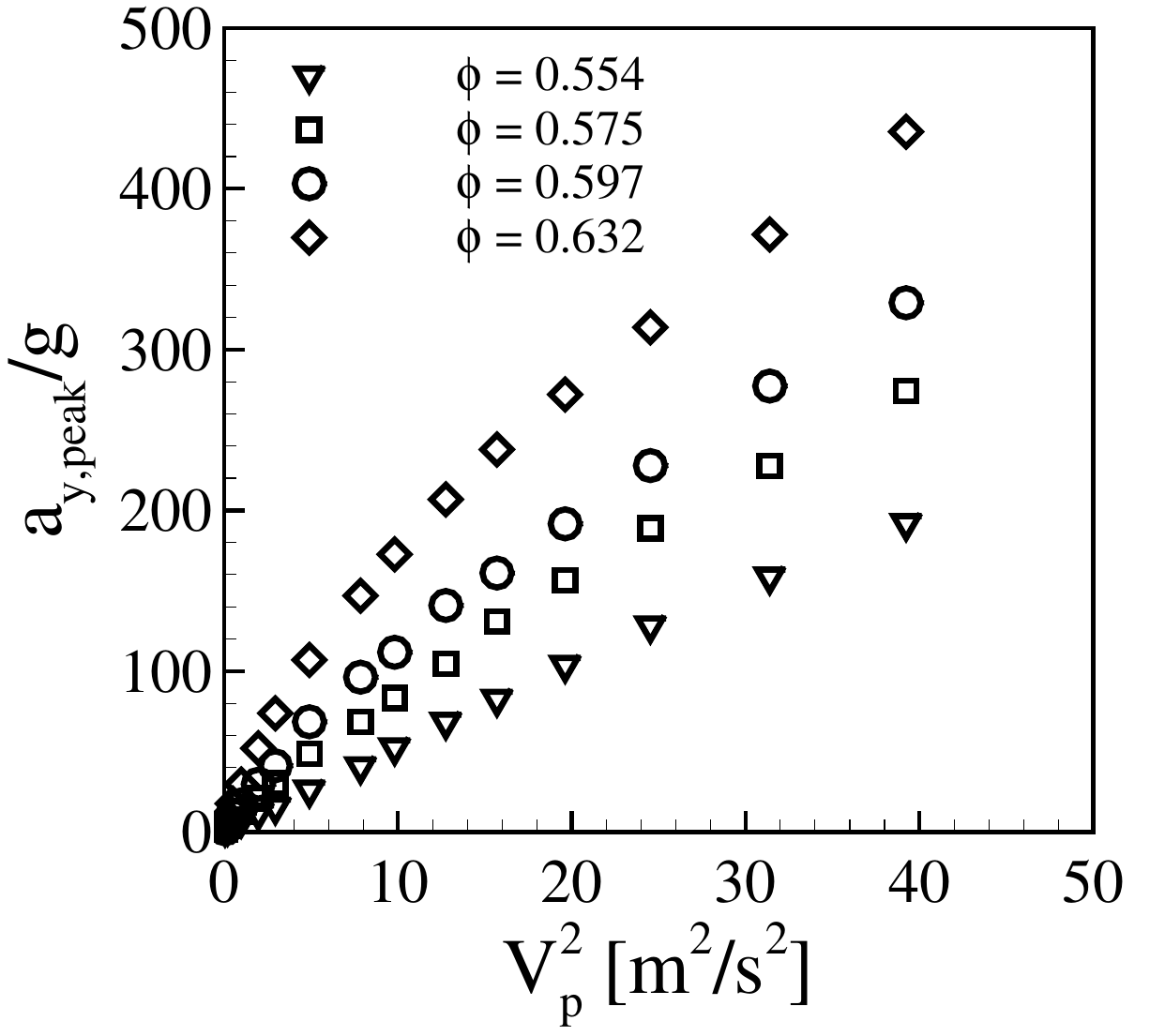}\\
				(b)
			\end{tabular}
		\end{minipage}
		\hfill
		\begin{minipage}{0.49\linewidth}
			\begin{tabular}{c}
				\\
				\includegraphics[width=0.8\linewidth]{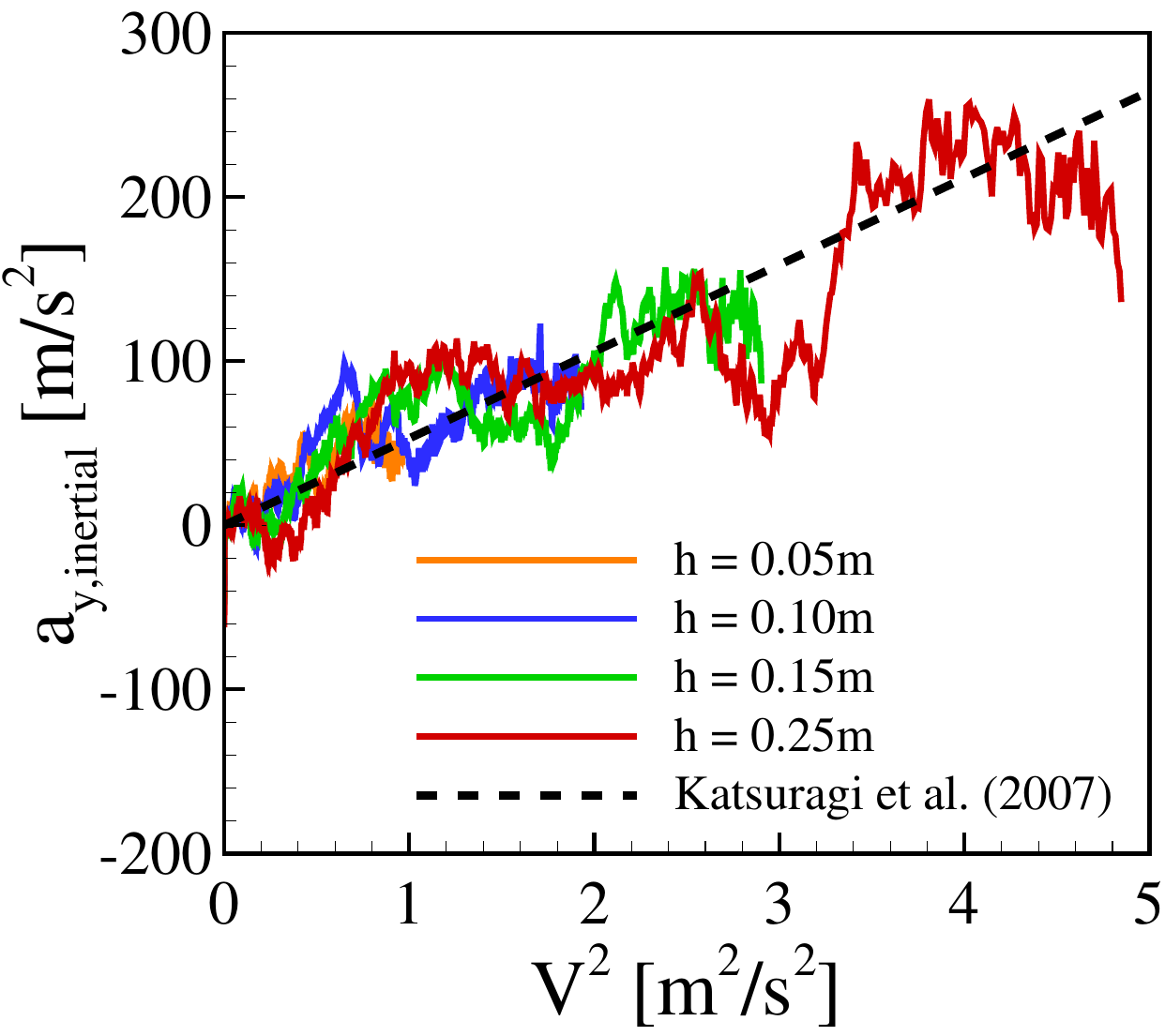}\\
				(c)
			\end{tabular}
		\end{minipage}
		\hfill		
	\end{center}
	\caption{Projectile dynamics. (a) Time evolution of the vertical component of the deceleration $a_y$ for different initial heights $h$. (b) Maximum values of the deceleration $a_{y,peak}$ as a function of $V_p^2$, for different packing fractions $\phi$. (c) Inertial component of the vertical deceleration, $a_{y,inertial}$, as a function of $V^2$ for different values of $h$. The dashed line corresponds to the model proposed by Katsuragi et al. \cite{Katsuragi} (with $\kappa$ = 37.6287 and $d_1$ = 0.0189). In Figures (a) and (c), the packing fraction was fixed to $\phi$ = 0.554.}
	\label{fig:drag}
\end{figure}

\begin{figure}[h!]
	\begin{center}
		\begin{minipage}{0.49\linewidth}
			\begin{tabular}{c}
				\\
				\includegraphics[width=0.8\linewidth]{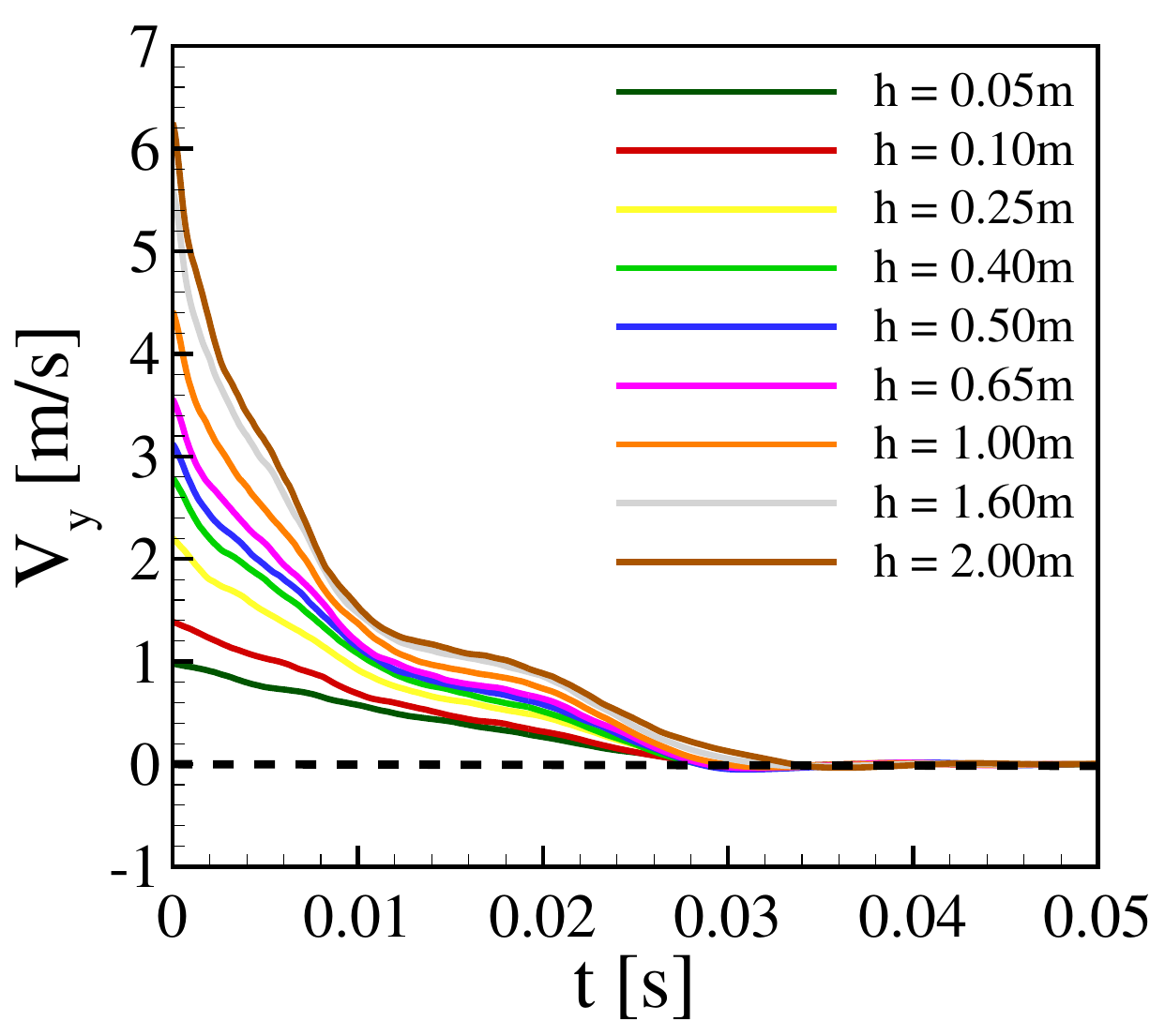}\\
				(a)
			\end{tabular}
		\end{minipage}
		\hfill
		\begin{minipage}{0.49\linewidth}
			\begin{tabular}{c}
				\\
				\includegraphics[width=0.8\linewidth]{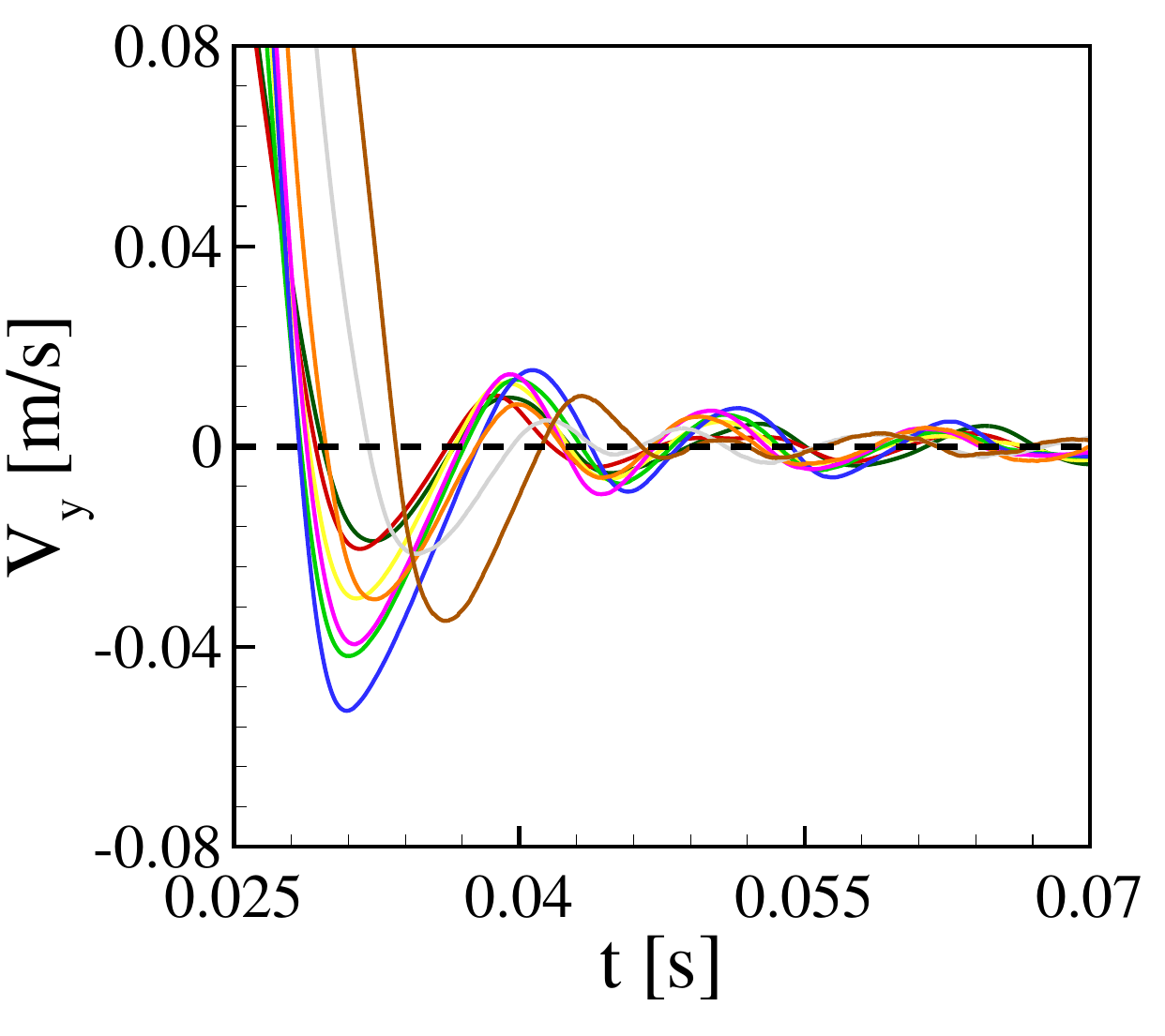}\\
				(b)
			\end{tabular}
		\end{minipage}
		\hfill
	\end{center}
	\caption{Time evolution of the vertical component of the projectile velocity $V_y$ for different values of $h$ for (a) the entire simulation and (b) zoomed in the region corresponding to the projectile rebound and final stop. The packing fraction was fixed to $\phi$ = 0.554.}
	\label{fig:drag2}
\end{figure}

We investigate now the projectile dynamics, in particular the accelerations experienced by the projectile and the time $t_c$ that it takes for reaching full stop. In the DEM simulations, positions, velocities and forces are computed for all objects at each time step, so that the time evolution of the projectile acceleration $\vec{a}$ can be obtained from the resultant force ($\vec{a}$ = $\vec{F}_p/m_p$), the drag force by subtracting the projectile weight from $\vec{F}_p$, and the stopping time by finding the instant when the projectile velocity $V$ reaches zero. In our analyses, the origin of time is the instant when the bottom of the projectile touches the granular bed, and $a_y$ is positive upwards (as $F_p$ in Eq. (\ref{Eq:intruder})).

Figure \ref{fig:drag}(a) shows the time evolution of the vertical component of the projectile deceleration $a_y$, normalized by $g$, for a fixed packing fraction ($\phi$ = 0.554) and different values of $h$, i.e., different energies available at the impact. We observe the features described by Goldman and Umbanhowar \cite{Goldman}: (i) a high peak just after the impact has taken place, with its magnitude increasing with $h$; (ii) the presence of strong fluctuations; (iii) a discontinuity of the deceleration by the end of the motion; and (iv) a slight inversion in the sign of $a_y$ before reaching full stop (see the Supplemental Material \cite{Supplemental} for a graphic showing the $a_y$ inversion and the full stop in detail, and a movie showing the projectile and grains during the impact). Besides reproducing the experimental findings of Goldman and Umbanhowar \cite{Goldman}, we can now inquire into aspects not previously investigated, such as the effect of the packing fraction and the mechanics of the projectile rebound.

Figure \ref{fig:drag}(b) presents the maximum values of the deceleration, $a_{y,peak}$, as a function of the square of the impact velocity $V_p$, for different packing fractions $\phi$. It is clear from Fig. \ref{fig:drag}(b) that the value of the peak increases not only with the available energy at the impact, but also with the packing fraction. In mechanical terms, the projectile deceleration is expected to increase with the bed compaction, since more grains are in contact as $\phi$ increases, hindering their motion and, consequently, that of the projectile. Although previous works showed the deceleration peak and its dependence on $h$, this is the first time that a parametric study on $\phi$ is presented, which corroborates the idea of higher $a_y$ for higher $\phi$ and, in part, the argument advanced by Goldman and Umbanhowar \cite{Goldman} that Eq. (\ref{Eq:intruder}) should depend on $\phi$ (and would be valid only close to a critical packing $\phi_{cps}$).

In order to inquire if our results agree with the model of Katsuragi et al. \cite{Katsuragi}, we evaluated the inertial term of the vertical deceleration, $a_{y,inertial}$. By considering $F_{drag}$ = $\xi V^2 + \kappa y$ in Eq. (\ref{Eq:intruder}), where $\xi V^2$ = $m_p\, a_{y,inertial}$ is the inertial term of the drag force (positive upwards), $a_{y,inertial}$ is given by Eq. (\ref{eq_ainertial}).

\begin{equation}
	a_{y,inertial} = a_y + g -(\kappa /m_p) y
	\label{eq_ainertial}
\end{equation}

\noindent Figure \ref{fig:drag}(c) shows $a_{y,inertial}$ as a function of the square of the instantaneous velocity of the projectile, $V^2$, for $\phi$ = 0.554 and different values of $h$, and also the model proposed by Katsuragi et al. \cite{Katsuragi} (dashed line). For all initial heights investigated (at $\phi$ = 0.554), the agreement with Katsuragi et al. \cite{Katsuragi} is good.

Figure \ref{fig:drag2} presents the time evolution of the vertical component of the projectile velocity, $V_y$, for different values of $h$, showing that the time $t_c$ to reach the full stop is independent of the available energy, in agreement with previous works \cite{Ciamarra, Katsuragi, Goldman, Seguin} (in Fig. \ref{fig:drag2}(a), $V_p$ $\geq$ 0.99 m/s). Furthermore, the stopping time $t_c$ scales well with the timescale $t_o$ $\sim$ $(\rho_p/ \rho_p)^{1/4} (D_p/(2g))^{1/2}$ proposed by Goldman and Umbanhowar \cite{Goldman}, which is $t_o$ = 0.0365 s in our case (very close to the values of $t_c$ in Fig. \ref{fig:drag2}), and the sign of $V_y$ changes just before the full stop (indicative of the final rebound).

\begin{figure}[h!]
	\begin{center}
		\includegraphics[width=.99\linewidth]{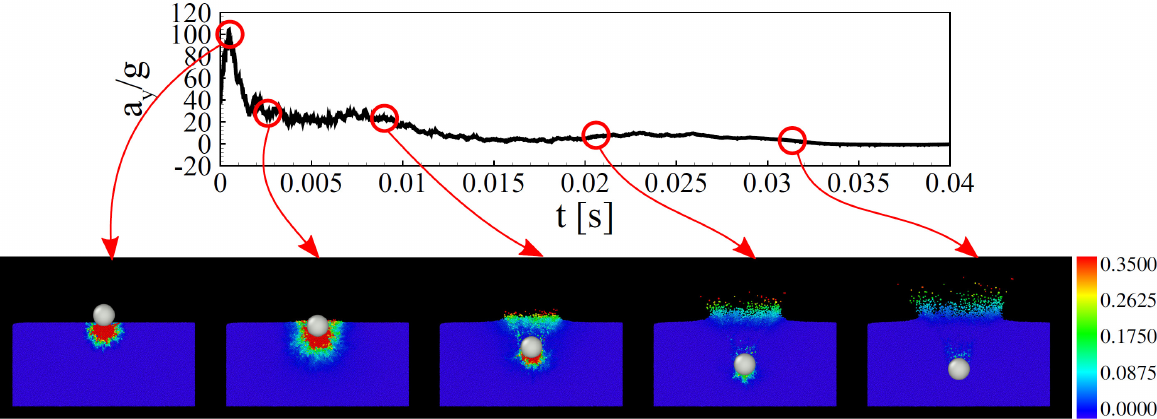}\\
	\end{center}
	\caption{Time evolution of the normalized deceleration in the vertical direction $a_y/g$, and snapshots of the granular temperature $\theta$ at the instants indicated in the $a_y/g$ graphic. The colorbar indicates the values of $\theta$ in m$^2$/s$^2$, and the figure corresponds to $\phi$ = 0.554 and $h$ = 1 m.}
	\label{fig:rebound_gratemp}
\end{figure}

Finally, we investigate the projectile rebounding. We begin by showing how the granular temperature $\theta$ evolves as the projectile penetrates into the bed. For that, we computed the granular temperature of the bed as in Eq. (\ref{eqngrantemp}),

\begin{equation}
	\theta(x,y,z,t) = \frac{1}{3} \vec{u_g'}^2 = \frac{1}{3} \left( u'^2 + v'^2 + w'^2 \right) \,,
	\label{eqngrantemp}
\end{equation}

\noindent where $\vec{u_g^{\prime}}$ is the instantaneous fluctuation velocity of each grain (its velocity relative to the ensemble of grains), and $u^{\prime}$, $v^{\prime}$ and $w^{\prime}$ are the $x$, $y$ and $z$ components of $\vec{u_g^{\prime}}$. Therefore, high values of $\theta$ indicate more agitation and a fluid-like behavior, while low values indicate less agitation and a solid-like behavior.

Figure \ref{fig:rebound_gratemp} shows the time evolution of the vertical deceleration of the projectile $a_y$ normalized by $g$, and snapshots of the granular temperature $\theta$ at some instants (indicated in the $a_y/g$ graphic), for $\phi$ = 0.554 and $h$ = 1 m. A movie showing the time evolution of $\theta$ during all the penetration process is available in the Supplemental Material \cite{Supplemental}. From both Fig. \ref{fig:rebound_gratemp} and the movie, we observe that initially ($t$ $\lessapprox$ 0.005 s) the region of higher granular temperatures is just below the projectile and, with its motion downwards, grains above the projectile reach higher values of $\theta$ at a later time (0.01 s $\lessapprox$ $t$ $\lessapprox$ 0.02 s), in particular the ejecta. By the end of its motion and before full stop (0.025 s $\lessapprox$ $t$ $\lessapprox$ 0.04 s), values of $\theta$ are considerable smaller, reaching zero below the projectile earlier than above it. This means that the region in front of the projectile (below it) is hardened (solid-like behavior) while that behind it (above the projectile) has still some mobility. Therefore, the rebound can be understood as a result of the faster de-fluidization on the front (bottom) than on the rear (top) of the projectile. We note that we have not inquired into shockwaves propagating from the impact point, toward the walls, and back to the projectile, which can play a role in the projectile rebound, as pointed out by Bourrier et al. \cite{Bourrier}. However, Bourrier et al. \cite{Bourrier} propose that the rebound of large projectiles is caused by the compaction of grains below the projectile, in agreement with our results (though we cannot assert that shockwaves are responsible for the rebound). 

A figure showing the vertical position $y$ of the projectile as a function of time for $t$ $>$ 0.025 s, and the displacement in the vertical direction $\Delta y_{rebound}$ for simulations with different values of $H$, is available in the Supplemental Material \cite{Supplemental}. For the latter, we noticed that, although the data oscillate considerably, it seems that $\Delta y_{rebound}$ increases with $H$ for small heights, and then reaches a plateau for $H$ $\approx$ 0.7 m (it remains, however, to be investigated further).

\subsection{\label{sec:friction} Frictionless grains}

\begin{figure}[h!]
	\begin{center}
		\begin{minipage}{0.49\linewidth}
			\begin{tabular}{c}
				\includegraphics[width=0.8\linewidth]{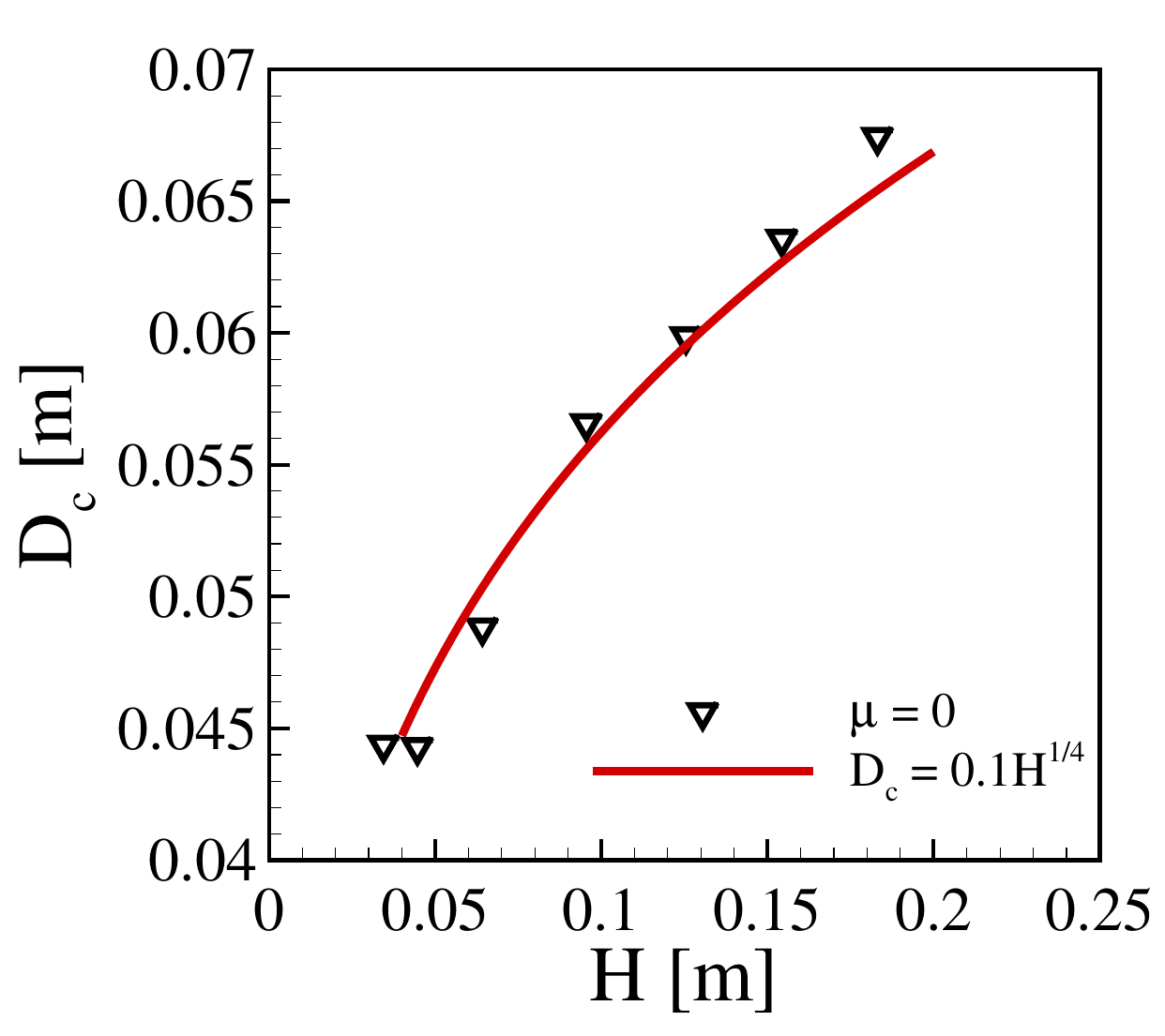}\\
				(a)
			\end{tabular}
		\end{minipage}
		\hfill
		\begin{minipage}{0.49\linewidth}
			\begin{tabular}{c}
				\includegraphics[width=0.8\linewidth]{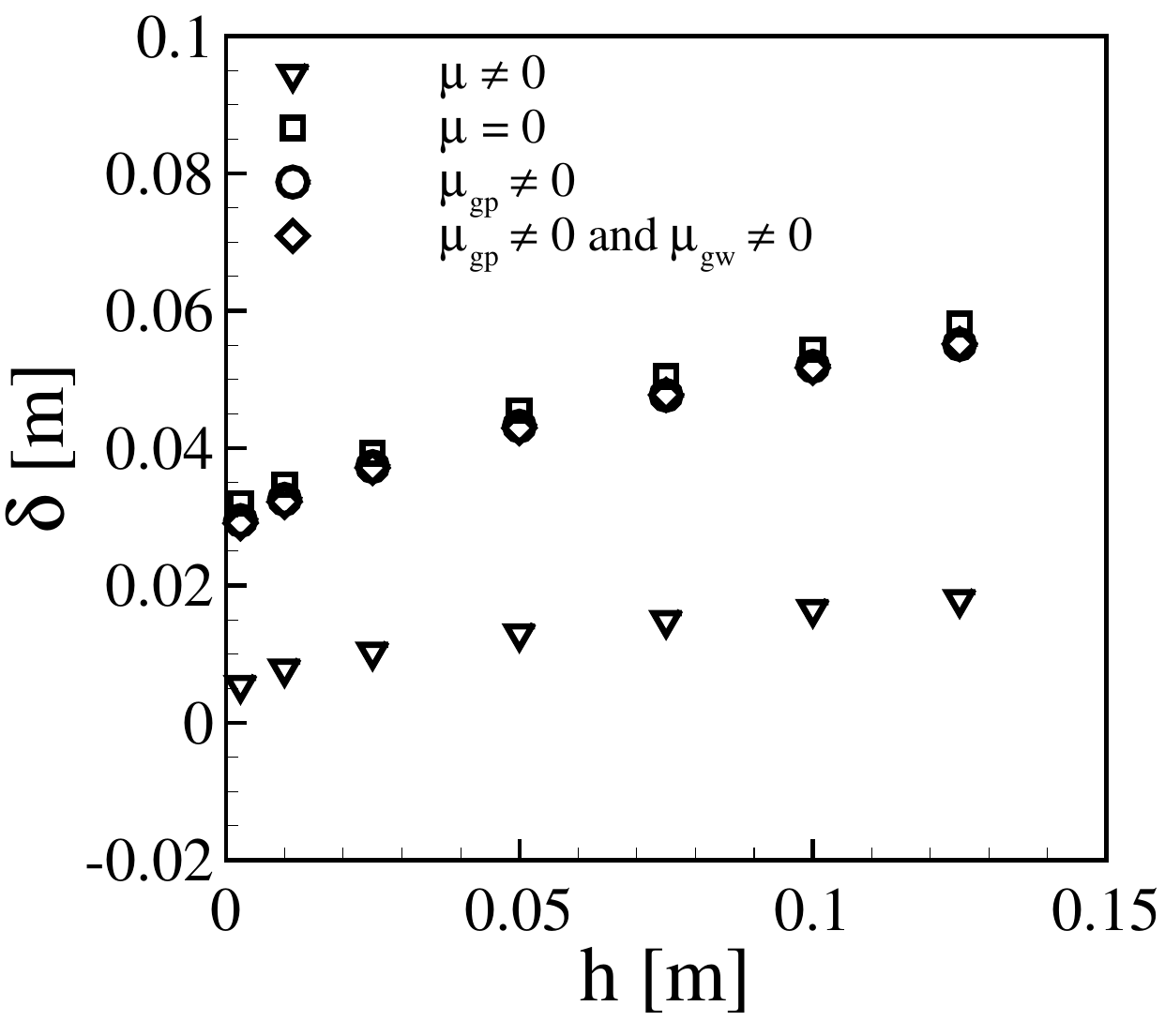}\\
				(b)
			\end{tabular}
		\end{minipage}
		\hfill
		\begin{minipage}{0.49\linewidth}
			\begin{tabular}{c}
				\\
				\includegraphics[width=0.8\linewidth]{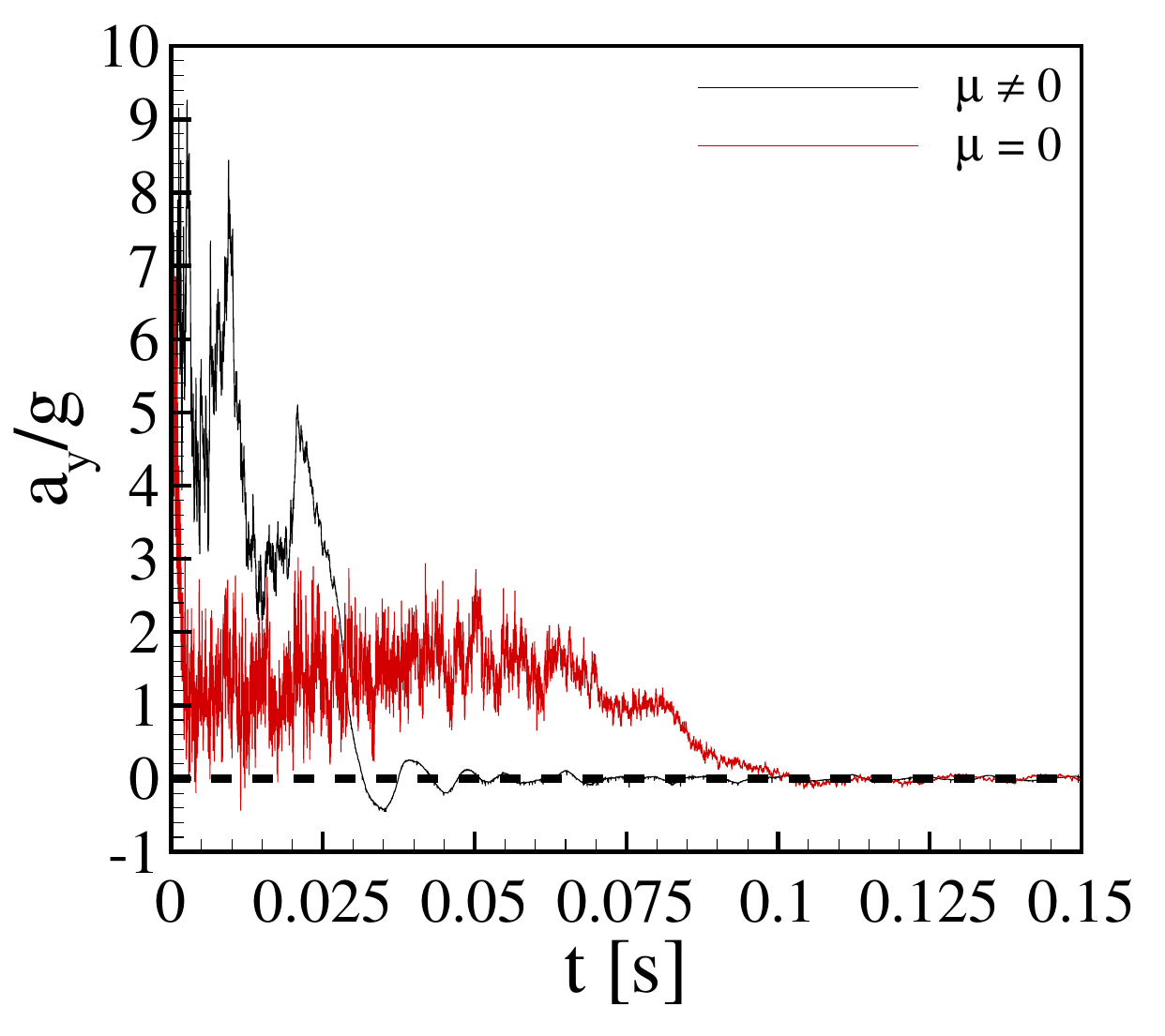}\\
				(c)
			\end{tabular}
		\end{minipage}
		\hfill
		\begin{minipage}{0.49\linewidth}
			\begin{tabular}{c}
				\\
				\includegraphics[width=0.8\linewidth]{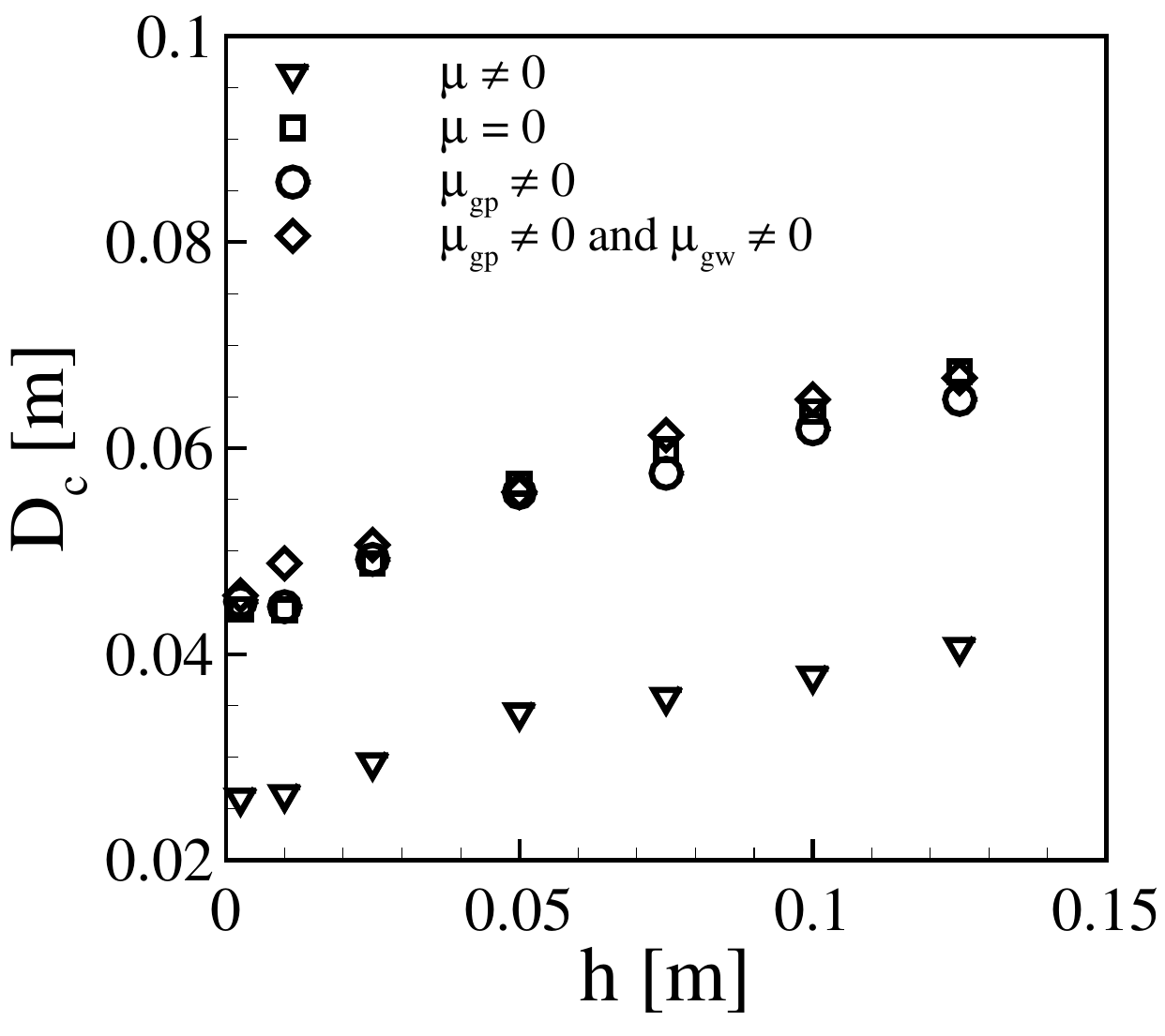}\\
				(d)
			\end{tabular}
		\end{minipage}
		\hfill
	\end{center}
	\caption{(a) Crater diameter $D_c$ as a function of the drop distance $H$ for simulations in the absence of friction (case $\mu$ $=$ 0). $\phi$ = 0.554 and the red line corresponds to a fitting with $H^{1/4}$. (b) Depth $\delta$ reached by the projectile as a function of the initial height $h$ for simulations with all friction coefficients as listed in Tab. \ref{tabcoefficients} (case $\mu$ $\neq$ 0), without any friction (case $\mu$ $=$ 0), with only the grain-projectile friction (case $\mu_{gp}$ $\neq$ 0), and with only the grain-grain equal to zero (case $\mu_{gp}$ $\neq$ 0 and $\mu_{gw}$ $\neq$ 0). (c) For the same cases of figure (b), $D_c$ as a function of $h$. (d) Time evolution of the normalized deceleration in the vertical direction $a_y/g$ for the cases with (black line) and without (red line) friction. In figure (d), $\phi$ = 0.554 when $\mu$ $\neq$ 0, and $h$ = 0.075 m.}
	\label{fig:frictionless}
\end{figure}

The role of friction in the projectile dynamics and cratering is still an open question, with previous works showing that the friction either promotes a strong energy dissipation \cite{Tsimring, Kondic} or does not affect the projectile penetration and stopping time \cite{Seguin} (those results are contradictory). In order to further investigate the role of friction, we carried out simulations that considered: (i) all friction coefficients as in Tab. \ref{tabcoefficients} (case $\mu$ $\neq$ 0); (ii) all friction coefficients equal to zero (case $\mu$ $=$ 0); (iii) only the grain-grain friction equal to zero (case $\mu_{gp}$ $\neq$ 0 and $\mu_{gw}$ $\neq$ 0); and (iv) the friction due to both the grain-grain and grain-wall contacts (but not grain-projectile) equal to zero (case $\mu_{gp}$ $\neq$ 0). In all these cases, whenever we indicate that $\mu_{gp}$, $\mu_{gw}$ or $\mu$ was turned to zero, the corresponding rolling frictions were also zero.

Figure \ref{fig:frictionless} presents the effects of the total or partial absence of friction on the crater diameter $D_c$ (Figs. \ref{fig:frictionless}(a) and \ref{fig:frictionless}(c)), penetration depth $\delta$ (Fig. \ref{fig:frictionless}(b)) and projectile deceleration in the vertical direction $a_y/g$ (Fig. \ref{fig:frictionless}(d)). We observe that $D_c$, $\delta$ and $a_y$ are highly affected by the absence of grain-grain friction, and that the presence/absence of the grain-wall and grain-projectile frictions have little effect on them. For the crater diameter, we observe that the $H^{1/4}$ scaling remains valid (Fig. \ref{fig:frictionless}(a)), but the magnitude of $D_c$ increases considerably in the absence of friction between grains, $D_c$ being roughly 50\% larger when $\mu$ = 0 (or at least $\mu_{gg}$ = 0) than when $\mu$ $\neq$ 0 (Fig. \ref{fig:frictionless}(c)). For the penetration depth, $\delta$ presents a variation with $h$ that is slighter when the grain-grain friction is present ($\mu$ $\neq$ 0), and has much smaller magnitudes than in the frictionless cases (roughly 75\% smaller, Fig. \ref{fig:frictionless}(b)). The vertical deceleration of the projectile, $a_y$, shows a different behavior in the absence of friction (Fig. \ref{fig:frictionless}(d)): it presents a smaller peak just after the impact, followed by a fast decrease to values that oscillate around $1.5g$, and finally a fast decrease to zero much after that of the frictional case ($t_c$ is much higher in the frictionless case, approximately by four times in Fig. \ref{fig:frictionless}(d)). Additional graphics for the granular temperature and projectile rebounding, and a movie of a projectile colliding with a frictionless bed are available in the Supplemental Material \cite{Supplemental}. Those graphics show an absence of rebound in the absence of friction.

In summary, our results show a strong influence of the grain-grain friction in both the morphology of craters and the projectile dynamics, which reflects the lower resistance to the projectile penetration when grain-grain friction is absent. The disagreement of our conclusions with those of Seguin et al. \cite{Seguin} is probably due to their highly confined 2D case. In our case, the simulations are 3D and wall effects are much less pronounced. Our results are important for deciding on the pertinence of the grain-grain friction and, therefore, for modeling and computing cratering in various scenarios.

\subsection{\label{sec:rotation} Rotating projectile (initial spin)}

\begin{figure}[h!]
	\begin{center}
		\begin{minipage}{0.49\linewidth}
			\begin{tabular}{c}
				\includegraphics[width=0.8\linewidth]{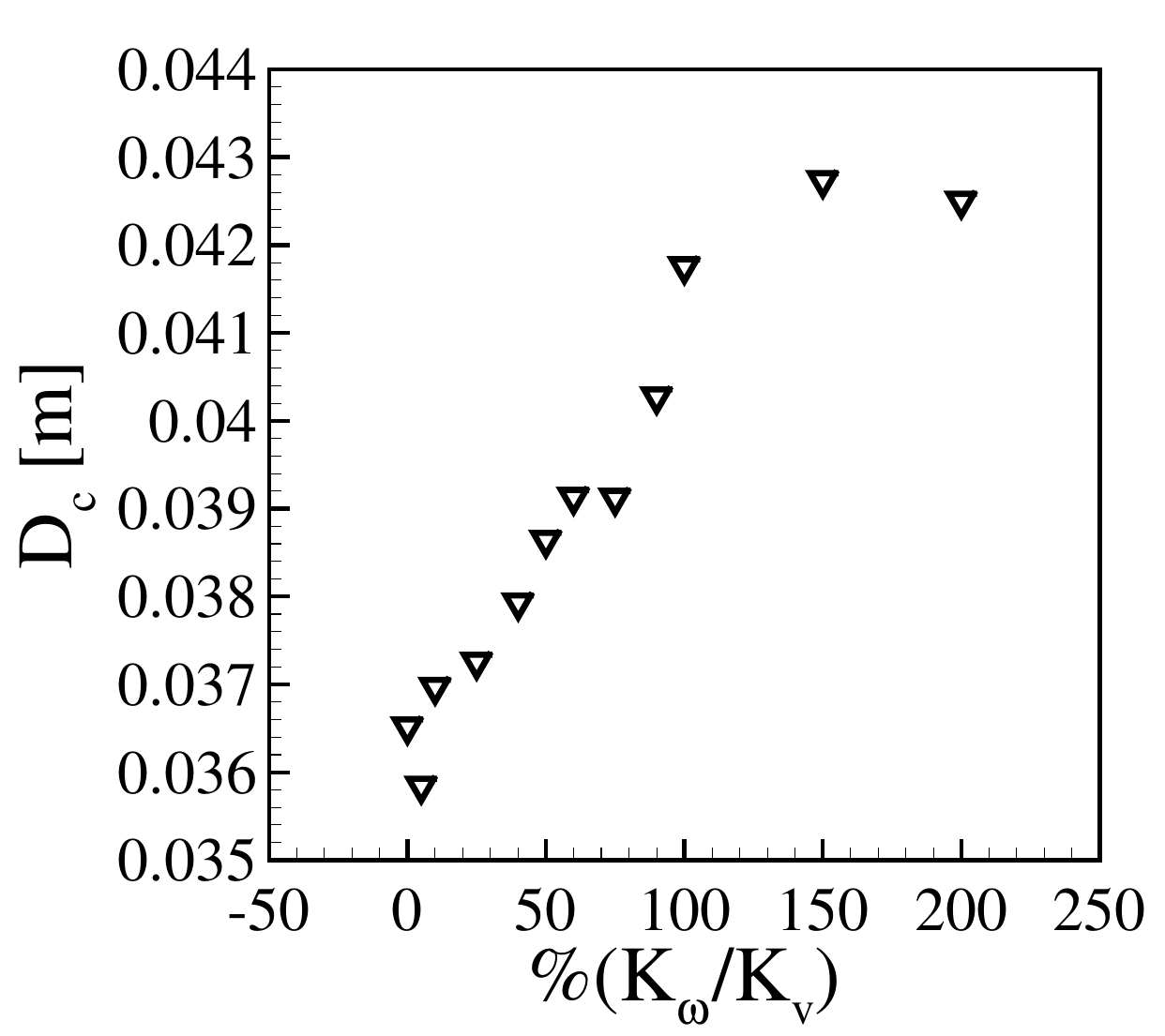}\\
				(a)
			\end{tabular}
		\end{minipage}
		\hfill
		\begin{minipage}{0.49\linewidth}
			\begin{tabular}{c}
				\includegraphics[width=0.8\linewidth]{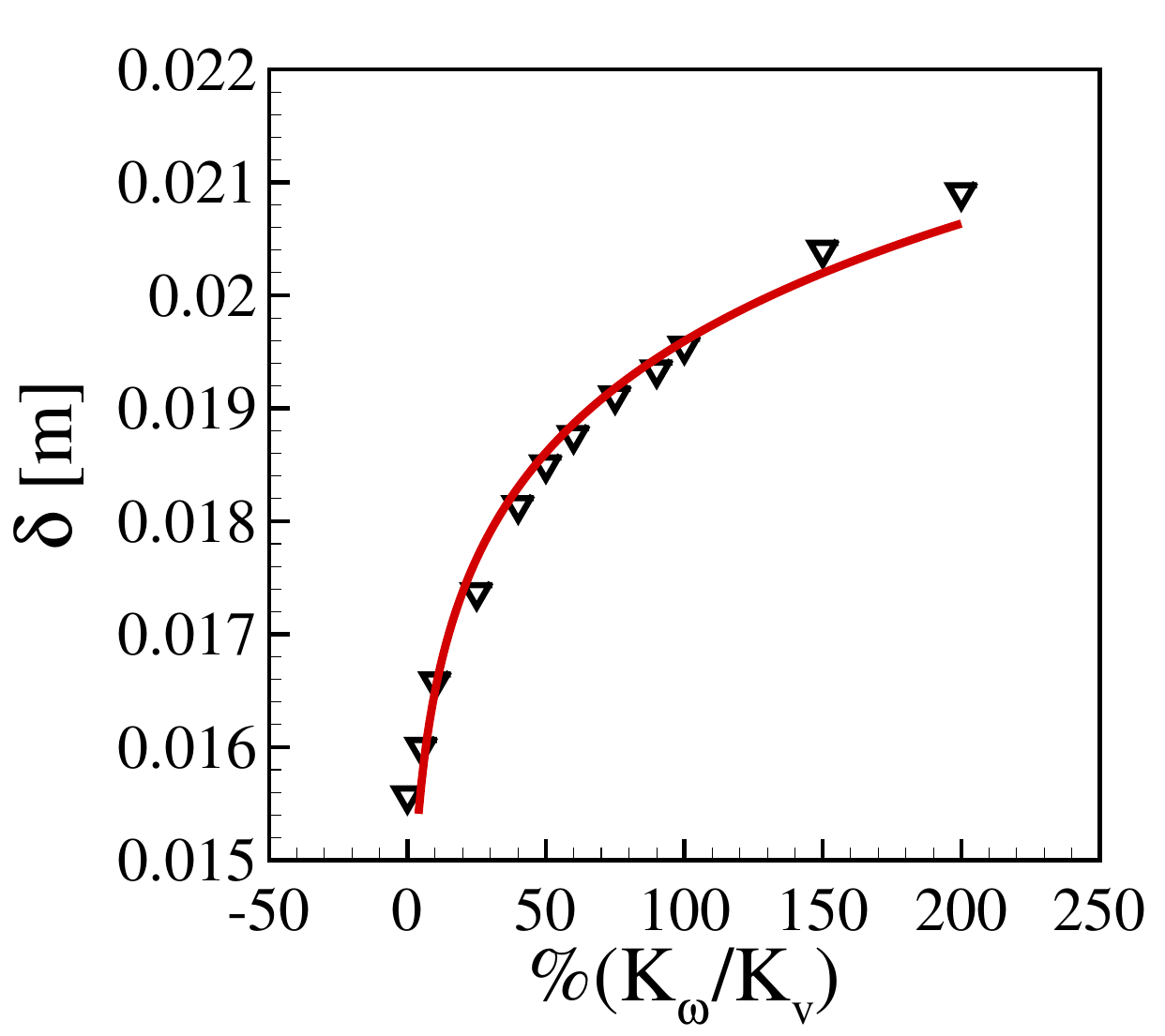}\\
				(b)
			\end{tabular}
		\end{minipage}
		\hfill
		\begin{minipage}{0.49\linewidth}
			\begin{tabular}{c}
				\\
				\includegraphics[width=0.8\linewidth]{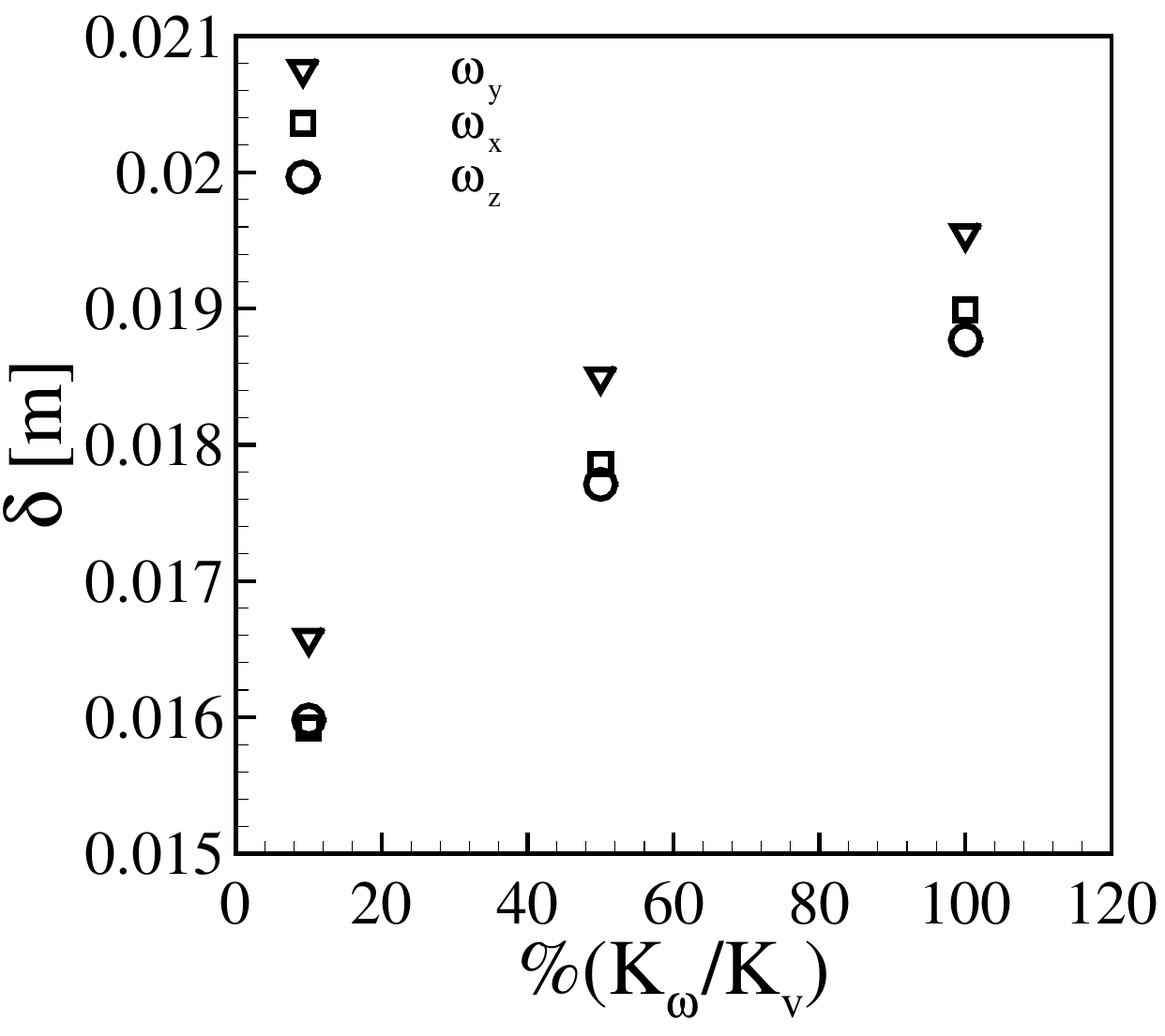}\\
				(c)
			\end{tabular}
		\end{minipage}
		\hfill
		\begin{minipage}{0.49\linewidth}
			\begin{tabular}{c}
				\\
				\includegraphics[width=0.8\linewidth]{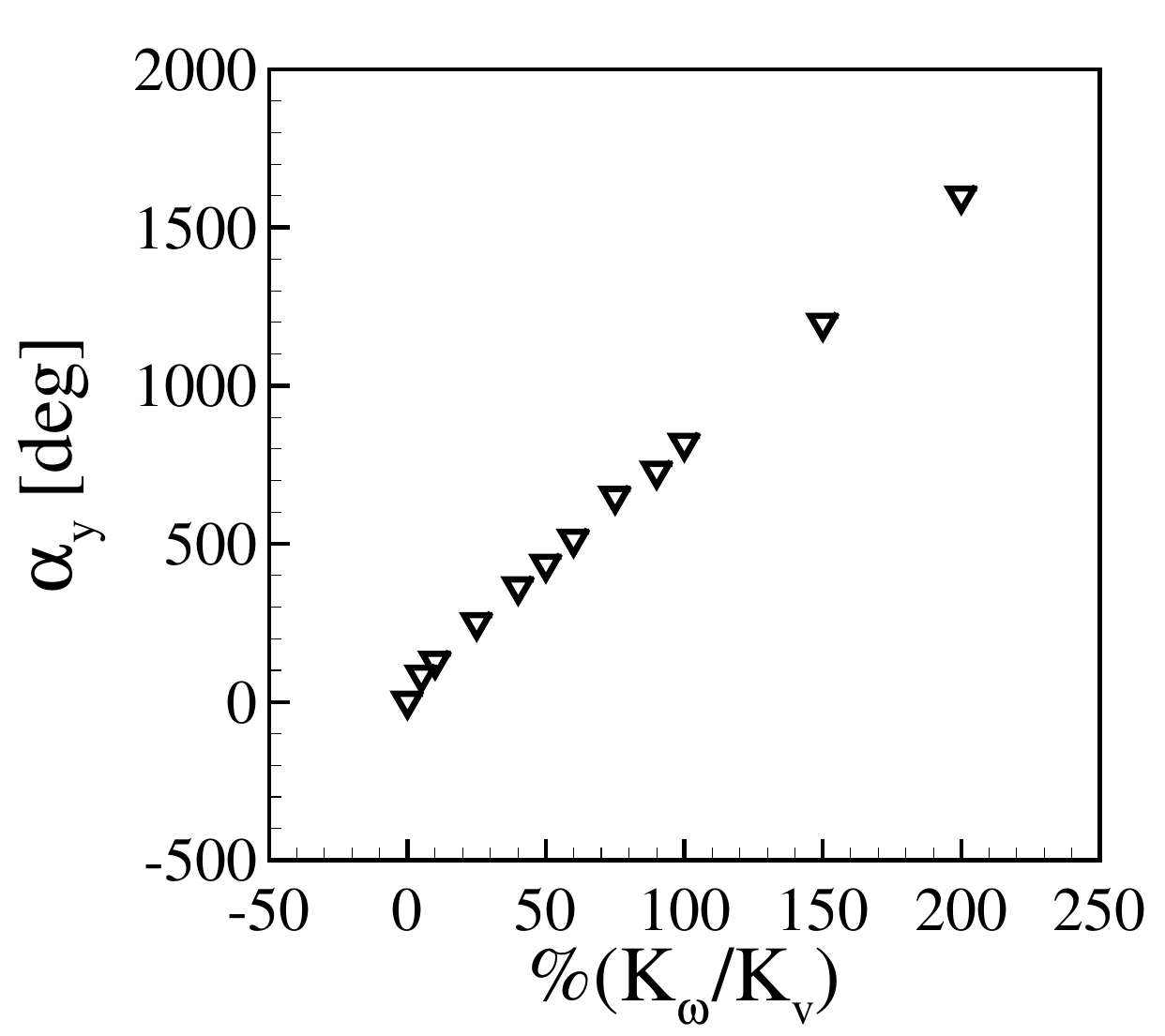}\\
				(d)
			\end{tabular}
		\end{minipage}
		\hfill
	\end{center}
	\caption{(a) Crater diameter $D_c$ as a function of the ratio of rotational to linear kinetic energies, $K_{\omega}/K_v$, in percentage, by considering only $\omega_y$. (b) Penetration depth $\delta$ as a function of $K_{\omega}/K_v$ by considering only $\omega_y$, and (c) for either $\omega_x$, $\omega_y$ or $\omega_z$ $\neq$ 0. (d) Total revolution angle (in degrees), $\alpha_y$, that the projectile effectuate after colliding with the bed as a function of $K_{\omega}/K_v$ (by considering only $\omega_y$). In figures (a) to (d), $\phi$ = 0.554 and $h$ = 0.1 m. In figure (b), the line corresponds to $\delta$ = 0.014 + $(K_{\omega}/K_v)^{0.075}$.}
	\label{fig:rotation}
\end{figure}

A question that has remained without investigation over the last decades, and that we scrutinize now, is the effect of the angular velocity of the projectile (rotational kinetic energy) on cratering. Our studies are motivated by the presence of spinning projectiles in natural and artificial processes, such as the fall of asteroids, weapon projectiles (spin imposed in order to stabilize their ballistic trajectory) and seeds (which acquire spin during their fall). To investigate this question, we carried out simulations where we imposed an initial angular velocity (initial spin) $\vec{\omega}$ to the projectile impacting the bed and computed the crater diameter, penetration depth and projectile dynamics for different $\vec{\omega}$ in terms of magnitude and direction. The angular velocity $\vec{\omega}$ was imposed only as an initial condition, the projectile being free to rotate or stop rotating in any direction after the impact has taken place (there was no constraint), so that it went to zero as the projectile finished penetrating the bed (excepting for frictionless solids, as explained next). In the following, we consider $\omega_x$, $\omega_y$ and $\omega_z$ the $x$, $y$ and $z$ components of $\vec{\omega}$, respectively, and $K_{v} = (1/2) m_p V_p^2$ and $K_{\omega} = (1/2) I |\vec{\omega}|^2$ the linear and rotational kinetic energies of the projectile, where $I = (2/5)m_p (D_p/2)^2$ is its moment of inertia.

\begin{figure}[h!]
	\begin{center}
		\includegraphics[width=0.8\linewidth]{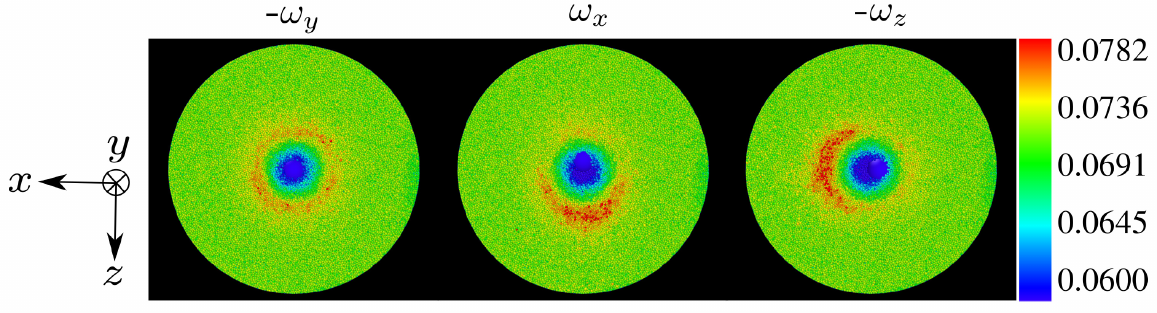}\\
	\end{center}
	\caption{Top views of final forms of craters resulting from projectiles with angular velocities in the $y$, $x$ and $z$ directions ($-\omega_y$, $\omega_x$ and $-\omega_z$, respectively). The colors correspond to $h_{bed}-y$ (the bed height measured from the bottom) and the values in the colorbar are in m. In this figure, $K_{\omega}/K_v$ = 1, $\phi$ = 0.554 and $h$ = 0.1 m.}
	\label{fig:rotation3}
\end{figure}

Figures \ref{fig:rotation}(a) and \ref{fig:rotation}(b) show, respectively, the crater diameter $D_c$ and penetration depth $\delta$ as functions of the ratio of rotational to linear kinetic energies, $K_{\omega}/K_v$, by varying only the $y$ component of the angular velocity, $\omega_y$ (both $\omega_x$ and $\omega_z$ were set to zero). We observe that both $D_c$ and $\delta$ vary with the rotation rate of the projectile, with $D_c$ and $\delta$ increasing by roughly 20 and 40\%, respectively, when $K_{\omega}/K_v$ varies from zero to two. Apparently, part of the rotational kinetic energy further agitates the bed, helping to dislodge more grains and excavate it (see the Supplemental Material \cite{Supplemental} for a graphic comparing the granular temperatures for rotating and non-rotating cases). In addition, we notice that, while a clear fitting cannot be found for $D_c$ (it seems to increase and then reach a plateau for $K_{\omega}/K_v$ $>$ 1, but we cannot assert it for the moment), $\delta$ follows a curve as in Eq. (\ref{proposed_correlation2}),

\begin{equation}
	\delta \,\sim\, \left( K_{\omega}/K_v \right)^n
	\label{proposed_correlation2}
\end{equation}

\noindent where $n$ = 0.075. In terms of the total rotation $\alpha_y$ that the projectile effectuate after colliding with the bed, Fig. \ref{fig:rotation}(d) shows a linear variation with $K_{\omega}/K_v$. In order to investigate the effect of the direction of $\vec{\omega}$ on the penetration depth $\delta$, we set either $\omega_x$, $\omega_y$ or $\omega_z$ to a nonzero value for $K_{\omega}/K_v$ equal to 10, 50 and 100\%. This is presented in Fig. \ref{fig:rotation}(c), which shows that in all cases $\delta$ follows the same trend with $K_{\omega}/K_v$, but with higher values for $\omega_y$ (by symmetry, $\omega_x$ and $\omega_z$ are equivalent). In the specific case of frictionless solids, $\delta$ reaches higher values and the projectile takes more time to stop rotating (when $\mu_{gp}$ $\neq$ 0) or even keeps rotating (when $\mu$ $=$ 0), though $\delta$ reaches a final value (see the Supplemental Material \cite{Supplemental} for a graphic of $\delta$ as a function of $K_{\omega}/K_v$ for frictionless grains). Concerning the general morphology of the crater, Fig. \ref{fig:rotation3} shows top views of final forms resulting from projectiles with angular velocities in the $y$, $x$ and $z$ directions ($-\omega_y$, $\omega_x$ and $-\omega_z$, respectively). We observe strong asymmetries when either $\omega_x$ or $\omega_z$ are nonzero, with grains accumulating (forming the corona) mostly in the direction of the tangential velocity, since they are partially excavated by the projectile rotation.

\begin{figure}[h!]
	\begin{center}
		\begin{minipage}{0.49\linewidth}
			\begin{tabular}{c}
				\includegraphics[width=0.80\linewidth]{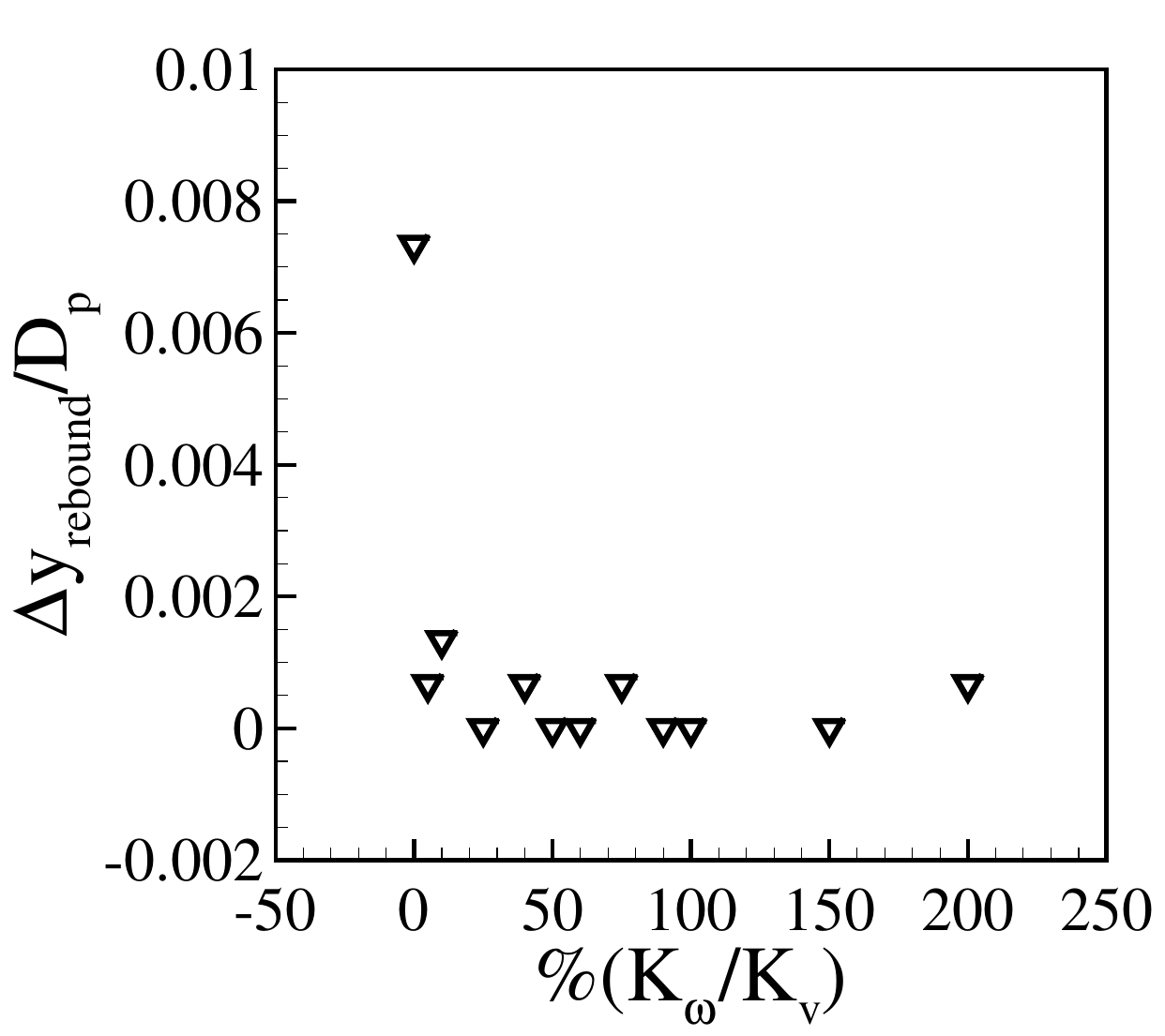}\\
				(a)
			\end{tabular}
		\end{minipage}
		\hfill
		\begin{minipage}{0.49\linewidth}
			\begin{tabular}{c}
				\includegraphics[width=0.80\linewidth]{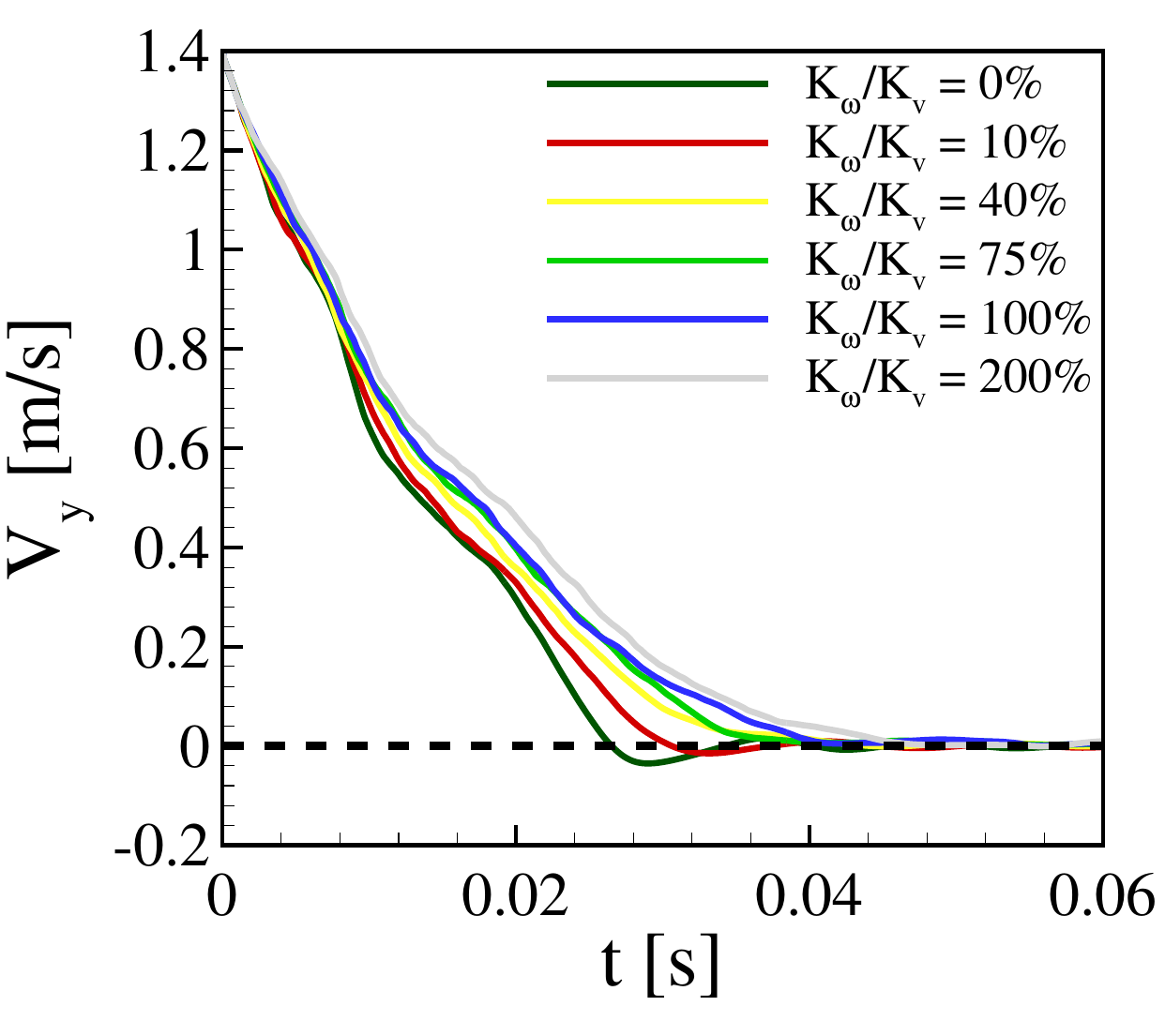}\\
				(b)
			\end{tabular}
		\end{minipage}
		\hfill\begin{minipage}{0.49\linewidth}
			\begin{tabular}{c}
				\includegraphics[width=0.80\linewidth]{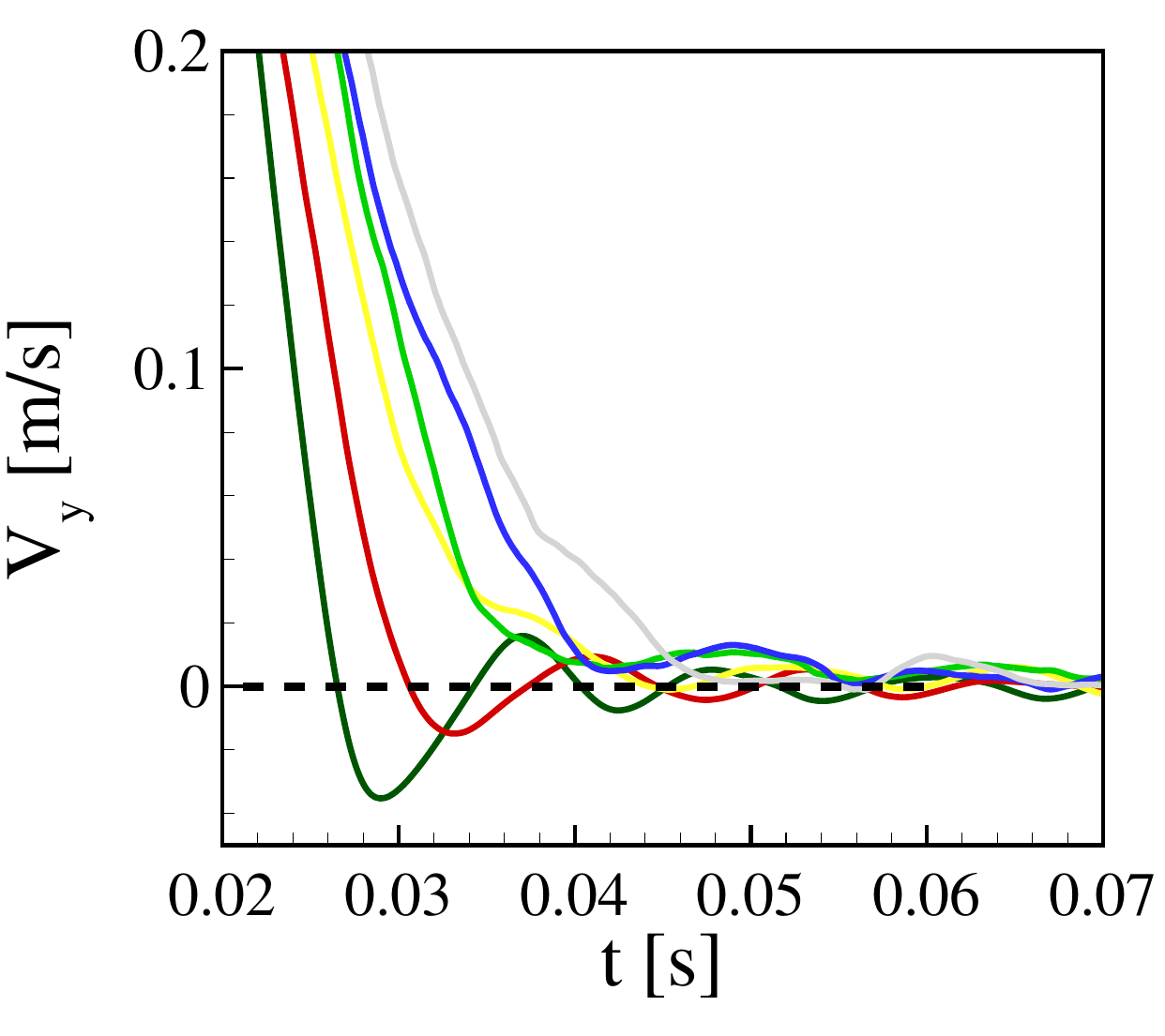}\\
				(c)
			\end{tabular}
		\end{minipage}
		\hfill
	\end{center}
	\caption{(a) Vertical displacement during the rebound $\Delta y_{rebound}$, normalized by $D_p$, as a function of the ratio of rotational to linear kinetic energies, $K_{\omega}/K_v$, in percentage. (b) Time evolution of the vertical component of the projectile velocity $V_y$ for different values of $K_{\omega}/K_v$ and the entire simulation, and (c) zoomed in the region corresponding to the projectile rebound and final stop. In figures (a) to (c), we consider only $\omega_y$, and $\phi$ = 0.554 and $h$ = 0.1 m.}
	\label{fig:rotation2}
\end{figure}

Finally, we measured the final rebound and time to reach the full stop for different values of $K_{\omega}/K_v$, which we present in Fig. \ref{fig:rotation2}, Fig. \ref{fig:rotation2}(a) showing the vertical displacement of the projectile during the rebound $\Delta y_{rebound}$ and Fig. \ref{fig:rotation2}(b) the time evolution of the vertical component of the projectile velocity $V_y$. We observe in Fig.  \ref{fig:rotation2}(a) that $\Delta y_{rebound}$ is approximately zero for rotating projectiles, indicating that in this case the rebound is suppressed even for small angular velocities ($K_{\omega}/K_v$ = 0.1), as can be see in detail in Fig. \ref{fig:rotation2}(c). While the rebound is suppressed, Figs. \ref{fig:rotation2}(b) and \ref{fig:rotation2}(c) show that the stopping time $t_c$ increases slightly with the angular velocity of the projectile.

To summarize, we investigated the impact of rotating projectiles with a granular bed, a common situation in nature, and analyzed the effects of rotation on cratering and projectile dynamics. We showed, for the first time, that both $D_c$ and $\delta$ vary with the angular velocity of the projectile and that the final rebound is suppresed by rotation. Additional graphics and tables are available in the Supplemental Material \cite{Supplemental}.

\section{CONCLUSIONS}
\label{sec:conclusions}

In this paper, we investigated numerically the formation of craters by an object impacting a granular bed, and concentrated our efforts into questions that were still open or to be investigated, such as the effects on cratering of the packing fraction of beds, solid friction of grains, and initial spin of projectiles. We found that the packing fraction $\phi$ does not affect the crater diameter $D_c$, both in terms of magnitude and functional relation with the drop distance $H$, while the depth $\delta$ reached by the projectile is highly influenced by $\phi$. By observing a lack of consensus in the literature, with diverging correlations for $\delta(H)$, and based on our results for different packing fractions, we proposed an \textit{ad hoc} scaling law that collapsed our data and indicates that some of the existing $\delta(H)$ correlations may be turned universal by considering $\phi$. For the projectile dynamics, we showed that it presents a high dependency on $\phi$, and explained the final rebound as the result of a faster de-fluidization on the front (bottom) than on the rear (top) of the projectile. We also showed that both the morphology of craters and the projectile dynamics are highly affected by the presence of fricionless grains (both  $D_c$ and $\delta$ increase with the absence of friction, $a_y$ oscillates around a constant value during great part of the penetration, and the projectile rebound is suppressed), evidencing the importance of grain-grain friction in models and computations. Finally, we revealed how  $D_c$ and $\delta$ increase with the initial spin (angular velocity) $\vec{\omega}$ of the projectile, and that the final rebound is suppressed by $\vec{\omega}$. Our results represent a new step toward understanding the mechanics of impact cratering in granular matter.

\section{\label{sec:Ack} ACKNOWLEDGMENTS}

The authors are grateful to FAPESP (Grant Nos. 2018/14981-7, 2019/20888-2 and 2020/04151-7) for the financial support provided.

\appendix
\section{Contact model}
\label{appendix}

The contact force between particles $i$ and $j$ , $\vec{F}_{c,ij}$, or between a particle $i$ and the wall, $\vec{F}_{c,iw}$, is usually decomposed into normal and tangential components, given by Eqs. (\ref{eqnFcn}) and (\ref{eqnFct}), respectively.

\begin{equation}
	F_{c,n} = \kappa_{n}\delta_{n} - \gamma_{n}\frac{d\delta_{n}}{dt}
	\label{eqnFcn}
\end{equation}

\begin{equation}
	F_{c,t} = \kappa_{t}\delta_{t} - \gamma_{t}\frac{d\delta_{t}}{dt}
	\label{eqnFct}
\end{equation}

The two terms in the RHS of Eq. (\ref{eqnFcn}) correspond to a repulsive force and a viscoelastic damping, and $\delta_{n} \geq 0$ is the normal displacement of two solids in contact. When two spherical particles are in contact, $\delta_{n}$ is given by:

\begin{equation}
	\delta_{n} = r_{i} + r_{j} - |\mathbf{x}_{i}-\mathbf{x}_{j}|
	\label{eqndeltan}
\end{equation}

\noindent where $r_{i}$ and $r_{j}$ are the radii of particles $i$ and $j$, and $\mathbf{x}_{i}$ and $\mathbf{x}_{j}$ the positions of their centers, respectively. For the contact between a spherical particle and a wall, $\delta_{n}$ is computed as the normal displacement between the center of the grain and the contact point. The two terms in the RHS of Eq. (\ref{eqnFct}) correspond to a shear force and a viscoelastic damping, and $\delta_{t}$ is the tangential displacement measured in the direction perpendicular to the plane of contact. $F_{c,t}$ is given by Eq. (\ref{eqnFct}) until it reaches $F_{c,t} = \mu F_{c,n}$, where $\mu$ is the microscopic coefficient of friction. From that moment, slip occurs and the tangential force becomes governed by the Coulomb's Law,

\begin{equation}
	F_{c,t} = \mu F_{c,n}
	\label{eqncoulomb}
\end{equation}

\noindent until the contact is finished. Coefficients $\kappa_{n}$, $\kappa_{t}$, $\gamma_{n}$, and $\gamma_{t}$ are functions of the displacements and grain properties. They are computed by Eqs. (\ref{eqnkappan}) to (\ref{eqntgamma}), based on the effective radius $r_{c}$, mass $m_{c}$, contact modulus $E_{c}$, and shear modulus $G_{c}$ (Eqs. (\ref{eqnrc}) to (\ref{eqnEc})) of particles $i$ and $j$ with, respectively, Young moduli $E_{i}$ and $E_{j}$ and Poisson's ratios $\nu_{i}$ and $\nu_{j}$,

\begin{equation}
	r_{c} = \frac{r_{i}r_{j}}{r_{i}+r_{j}}
	\label{eqnrc}
\end{equation}

\begin{equation}
	m_{c} = \frac{m_{i}m_{j}}{m_{i}+m_{j}},
	\label{eqnmc}
\end{equation}

\begin{equation}
	E_{c} = \Bigg(\frac{1 - \nu_{i}^{2}}{E_{i}} + \frac{1 - \nu_{j}^{2}}{E_{j}}\Bigg)^{-1}
	\label{eqnEc}
\end{equation}

\begin{equation}
	G_{c} = \Bigg[\frac{2(2 - \nu_{i})(1 + \nu_{i})}{E_{i}} + \frac{2(2 - \nu_{j})(1 + \nu_{j})}{E_{j}}\Bigg]^{-1}
	\label{eqnGc}
\end{equation}

\begin{equation}
	\kappa_n = \frac{4}{3} E_c \sqrt{R_c\delta_n}
	\label{eqnkappan}
\end{equation}

\begin{equation}
	\kappa_t = 8 G_c \sqrt{R_c\delta_n}
	\label{eqnkappat}
\end{equation}

\begin{equation}
	\gamma_n = -2\sqrt{\frac{5}{6}} \beta \sqrt{2E_c \sqrt{R_c\delta_n} m_c}
	\label{eqngamman}
\end{equation}

\begin{equation}
	\gamma_t = -2\sqrt{\frac{5}{6}} \beta \sqrt{8 G_c \sqrt{R_c\delta_n} m_c}
	\label{eqntgamma}
\end{equation}

\noindent where $\beta$ is a damping coefficient based on the restitution coefficient $\epsilon$, computed as in Eq. (\ref{eqnbeta}).

\begin{equation}
	\beta = \frac{\ln(\epsilon)}{\sqrt{\ln^{2}(\epsilon) + \pi^2}}
	\label{eqnbeta}
\end{equation}

For a particle of radius $r$, contact torques are computed as the sum of the torques due to $F_{c,t}$ and rolling friction, for all its contacts. For $\vec{T}_r$ representing the torque caused by rolling friction, contact torques can thus be summarized as in Eq. (\ref{eqntorques}):

\begin{equation}
	\vec{T}_{c} = \sum \left( r F_{c,t}\vec{n} \times \vec{t} + \vec{T}_{r}\right)
	\label{eqntorques}
\end{equation}

\noindent where $\vec{n}$ and $\vec{t}$ are unit vectors in the normal and tangential directions, respectively. $\vec{T}_r$ can be modeled as having spring and damping components, but Derakhshani et al. \cite{Derakhshani} showed that the damping component is negligible for DEM computations. Therefore,

\begin{equation}
	\vec{T}_{r} = -k_r \Delta \theta_r \vec{n} \times \vec{t}
	\label{eqntorquerolling}
\end{equation}

\noindent where $\theta_r$ is the incremental rolling at the considered contact and $k_r$ is the rolling stiffness, given by  Eq. (\ref{eqnrollingstiff}). 

\begin{equation}
	k_r = \mu_r R_c \frac{F_{c,n}}{\theta_r^m}
	\label{eqnrollingstiff}
\end{equation}

In Eq. (\ref{eqnrollingstiff}), $\theta_r^m$ is the angle for incipient rolling and $\mu_r$ is the coefficient of rolling resistance.

\bibliography{references}

\end{document}